\documentclass[useAMS,usenatbib]{mn2e}                                  
\usepackage{graphicx, natbib, amssymb, color, amsmath}

\title[MultiDark-Clusters Mock Light-Cones]{MultiDark-Clusters: Galaxy Cluster Mock Light-Cones, eROSITA and the Cluster Power Spectrum}

\author[Zandanel et al.]
{Fabio Zandanel$^{1,*}$, 
 Mattia Fornasa$^{1}$,
 Francisco Prada$^{2}$,
 Thomas H. Reiprich$^{3}$,  
\newauthor
 Florian Pacaud$^{3}$, 
 Anatoly Klypin$^{4}$\\
$^1$ GRAPPA Institute, University of Amsterdam, 1098XH Amsterdam, The Netherlands\\
$^2$ Instituto de Astrof\'{\i}sica de Andaluc\'{\i}a, CSIC, Apdo. Correos 3004, E-18080 Granada, Spain\\
$^3$ Argelander-Institut f{\"u}r Astronomie, Universit{\"a}t Bonn, Auf dem H{\"u}gel 71, 53121 Bonn, Germany\\
$^4$ Department of Astronomy, New Mexico State University, Las Cruces, NM 88003-0001, USA \\
}

\begin{document}

\date{Accepted 2018 July 13. Received 2018 July 13; in original from 2018 April 19.}

\pagerange{\pageref{firstpage}--\pageref{lastpage}} \pubyear{2018}

\maketitle		

\label{firstpage}

%%%%%%%%%%%%%%%%%%%%%%%%%%%%%%%%%%%%%%%%%%%%%%%%%%%%%%%%%%%%%%%%%%%
%%%%%%%%%%%%%%%%%%%%%%%%%%%%%%%%%%%%%%%%%%%%%%%%%%%%%%%%%%%%%%%%%%%
\begin{abstract}
Cosmological simulations are fundamental tools to study structure formation and the astrophysics of evolving structures, in particular clusters of galaxies. While hydrodynamical simulations cannot sample efficiently large volumes and explore different cosmologies at the same time, N-body simulations lack the baryonic physics that is crucial to determine the observed properties of clusters. One solution is to use \mbox{(semi-)}analytical models to implement the needed baryonic physics. In this way, we can generate the many mock universes  that will be required to fully exploit future large sky surveys, such as that from the upcoming eROSITA X-ray telescope. We developed a phenomenological model based on observations of clusters to implement gas density and temperature information on the dark-matter-only halos of the MultiDark simulations. We generate several full-sky mock light-cones of clusters for the WMAP and \emph{Planck} cosmologies, adopting different parameters in our phenomenological model of the intra-cluster medium. For one of these simulations and models, we also generate 100 light-cones corresponding to 100 random observers and explore the variance among them in several quantities. In this first paper on MultiDark mock galaxy cluster light-cones, we focus on presenting our methodology and discuss predictions for eROSITA, in particular, exploring the potential of angular power spectrum analyses of its detected (and undetected) cluster population to study X-ray scaling relations, the intra-cluster medium, and the composition of the cosmic X-ray background. We make publicly available on-line more than $400$~GB of light-cones, which include the expected eROSITA count rate, on Skies \& Universes (http://www.skiesanduniverses.org).
\end{abstract}


%%%%%%%%%%%%%%%%%%%%%%%%%%%%%%%%%%%%%%%%%%%%%%%%%%%%%%%%%%%%%%%%%%%
%%%%%%%%%%%%%%%%%%%%%%%%%%%%%%%%%%%%%%%%%%%%%%%%%%%%%%%%%%%%%%%%%%%
\section{Introduction}
\label{sec:1}

\begingroup
\let\thefootnote\relax\footnotetext{* fabio.zandanel@gmail.com}
\endgroup
 
Clusters of galaxies represent the latest stage of structure formation in our Universe (e.g., \citealp{2005RvMP...77..207V,2011ARA&A..49..409A}). They are beautifully complex objects that encompass many astrophysical phenomena. Cluster mergers are the most energetically violent events that we know, releasing energies of the order of the final gas binding energy up to
about $10^{64}$~erg, and the non-thermal phenomena connected to these processes, such as turbulence and particle acceleration, sometime give rise to spectacular diffuse synchrotron radio emission on Mpc scales (e.g., \citealp{2014IJMPD..2330007B}). 
Clusters are permeated by a thermal gas with keV temperatures called the intra-cluster medium (ICM), which can be traced with X-ray observations via its bremsstrahlung and line emission (e.g., \citealp{1988xrec.book.....S}) and with its thermal Sunyaev-Zel'dovich (SZ) signal (e.g., \citealp{2002ARA&A..40..643C}). These thermal properties of clusters of galaxies, as well as weak-lensing observations (e.g, \citealp{2010CQGra..27w3001B}), can be used to trace their dark matter content and, therefore, their total mass: a fundamental ingredient to study cosmology. 

With the goal to both exploring and fully exploiting the potential of current and upcoming samples of clusters of galaxies, in particular from SZ and X-ray surveys, we embarked on an ambitious project of generating several hundreds of clusters mock light-cones starting from dark-matter-only N-body simulations. With \emph{light-cone} we mean a collection of clusters properly distributed in the sky according to a certain prescription and starting from the halo catalogues of a N-body simulation: a mock universe containing only galaxy clusters, i.e., halos above a certain mass. One final objective is to be able to directly estimate cosmic variance at the scale of galaxy clusters, and how this translates into uncertainties on various statistical properties expected to be measured (e.g., scaling relations, mass function, power spectrum), rather than rely only on analytical approaches (e.g., \citealp{2011A&A...536A..95V}). Note that in the following, we will call \emph{sample variance} the variance dominated by differences in the considered samples as, for example, when analysing small patches of the sky as in Section~\ref{sec:5.1}, and we will call \emph{cosmic variance} that due to the limitation of observing only this one Universe. In this work, we present our starting efforts in this direction. In particular, we introduce the methodology used to populate N-body simulations with baryonic physics, and focus on two applications: predictions for cluster population of the upcoming eROSITA X-ray satellite\footnote{http://www.mpe.mpg.de/eROSITA} \citep{2010SPIE.7732E..0UP,2012arXiv1209.3114M,2012MNRAS.422...44P,2014A&A...567A..65B,2018arXiv180608652C}, and for its power spectrum. We argue that the latter is a particularly promising way not only to put constraints on cosmology (e.g., \citealp{2001A&A...368...86S,2005MNRAS.362.1225S,2011MNRAS.413..386B,2012MNRAS.422...44P}) but also to study X-ray scaling relations and the gas profile in the outskirts of clusters (e.g., \citealp{2017MNRAS.466.3035K}; \citealp{2018MNRAS.473.4653K}; \citealp{2010ApJ...725.1452S}; \citealp{2014A&A...568A..57H}), and to study the composition of the cosmic X-ray background (e.g., \citealp{2012MNRAS.427..651C,2017MNRAS.466.3035K}).

The advantage and complementarity of using N-body over hydrodynamical simulations is two fold in this context. First, for the same of box dimension, resolution and number of particles, an N-body simulations is cheaper to run than a hydrodynamical one. In this work, we use the MultiDark suite of N-body simulations \citep{2016MNRAS.457.4340K} that, in fact, is composed of several runs with different cosmological conditions, box sizes and resolutions. As of today, it is computationally prohibitive to have a similar set, composed of several different simulations, of hydrodynamical runs.
Secondly, we use an analytical model, either constructed from first principles (e.g., \citealp{2010ApJ...720.1038B,2010ApJ...725.1452S}) or phenomenologically from available observations (e.g., \citealp{2012MNRAS.425.2244B,2013arXiv1311.4793Z}), to implement the needed baryonic physics (i.e., gas density and temperature) for the dark-matter halos. Despite the great simplifications (e.g., spherical symmetry of the clusters), this allows us to easily change the recipe of the model and test several scenarios without the need of running from scratch new simulations. Needless to say, the advantages of this approach are not suited for all purposes. For example, it is well suited for statistical studies on the cluster population, as in the present paper, but not for studying cluster-to-cluster differences and physical peculiarities. In this sense, hydrodynamical simulations, such as MUSIC \citep{2013MNRAS.429..323S}, Illustris \citep{2014MNRAS.444.1518V}, Magneticum \citep{2016MNRAS.463.1797D}, and BAHAMS \citep{2017arXiv171202411M}, remain indispensable tools.

This paper is structured as follows. In Section~\ref{sec:2}, we explain the steps from the selection of the N-body simulations to the final galaxy clusters mock light-cones, and in Section~\ref{sec:3} we validate our phenomenological modelling for the gas density and temperature against current X-ray and SZ observations. We make predictions tailored for the outcome of the eROSITA full sky survey and provide few examples of the potential of our light-cones in Section~\ref{sec:4}. We then turn to the study of the power spectrum of the X-ray emission of clusters of galaxies in Section~\ref{sec:5}: we compare the results from our light-cones to current observations and make predictions for eROSITA. We discuss some limitations of the present light-cones and plans for the future in Section~\ref{sec:6}. Finally, in Section~\ref{sec:7} we present the summary and conclusions. In Appendix~\ref{app:A}, we describe in detail the produced mock light-cones that we make publicly available on-line at Skies \& Universes \citep{2017arXiv171101453K}.\footnote{http://www.skiesanduniverses.org}

Note that the cosmological parameters adopted through the paper correspond to those adopted in the MultiDark simulations (see Table~1) and are not necessarily always the same. The cluster masses $M_{\Delta}$ and radii $R_{\Delta}$, unless specified otherwise, are defined with respect to a density that is $\Delta=500$ times the \emph{critical} density of the Universe at the cluster redshift. Note also that, in the units of measure, we always explicit the dependence on the Hubble constant $H_0 = 100\, h_{X}$~km~s$^{-1}$~Mpc$^{-1}$ where $h_{X}$ can be equal to $h=1$ or to $h_\mathrm{cosmo}$ which, in turn, can be $h_{70} = 0.7$ (for the WMAP-7 cosmology) or $h_{68}=0.678$ (for the \emph{Planck} cosmology) depending on the used MultiDark simulation (see Table~1).

%%%%%%%%%%%%%%%%%%%%%%%%%%%%%%%%%%%%%%%%%%%%%%%%%%%%%%%%%%%%%%%%%%%
%%%%%%%%%%%%%%%%%%%%%%%%%%%%%%%%%%%%%%%%%%%%%%%%%%%%%%%%%%%%%%%%%%%
\section{Method for the construction of the light-cone catalogues}
\label{sec:2}

The process of generating galaxy cluster mock light-cone catalogues comprises two main steps. The first is the generation of the light-cone halo catalogue itself from the dark-matter-only N-body simulations, and the second is the implementation of the phenomenological models for the gas density and temperature for the simulated halos in the light-cones. We describe in detail these two steps in the following subsections.

%%%%%%%%%%%%%%%%%%%%%%%%%%%%%%%%%%%%
%%%%%%%%%%%%%%%%%%%%%%%%%%%%%%%%%%%%
\subsection{MultiDark simulations and light-cones}
\label{sec:2.1}

\begin{table*}
\begin{center}
\caption{Numerical and cosmological parameters for the MultiDark simulations used in this paper, both run with the GADGET-2 code.
Columns show the simulation identifier, size of the simulated box, number of particles, particle mass, the Plummer equivalent gravitational 
softening length $\epsilon$, the adopted values for $\Omega_{\rm{Matter}} (\Omega_{\rm{M}}),\, \Omega_{\rm{Baryon}} (\Omega_{\rm{B}}),\, \Omega_{\Lambda}$, the clustering 
at $8~h^{-1}\,{\rm Mpc}$, $\sigma_8$, the spectral index $n_s$, and the Hubble constant at present day. See \protect\cite{2016MNRAS.457.4340K} for more details.}
\begin{tabular}{ l c | c | c | c | c | c | c | c | c l c | c }
\hline  
Simulation & box [$h^{-1}\,{\rm Gpc}$] &  particles  & $m_p$ [$h^{-1}\,M_\odot$] & $\epsilon$ [$h^{-1}\,{\rm kpc}$] &
$\Omega_M$ & $\Omega_B$ & $\Omega_{\Lambda}$ & $\sigma_8$ & $n_s$ & $H_0 [${\rm km/s/Mpc}$]$
\tabularnewline
  \hline   
BigMD27    & $2.5$    &  $3840^3$   &  $2.1 \times 10^{10}$     & $10.0$ &
$0.270$  & $0.047$   &  $0.730$  &  $0.820$  &  $0.95$  &  $70.0$
\tabularnewline
BigMDPL  & $2.5$    &  $3840^3$   &  $2.4 \times 10^{10}$  & $10.0$ & 
$0.307$  & $0.048$   &  $0.693$  &  $0.829$  &  $0.96$  &  $67.8$
\tabularnewline
\hline
\end{tabular}
\end{center}
\label{tab:simtable}
\end{table*}
  
In this work we use two of the MultiDark suite simulations dubbed BigMD27 and BigMDPL which refers to WMAP-7 \citep{2011ApJS..192...18K} and \emph{Planck} year-1 cosmologies \citep{2014A&A...571A..16P}, respectively. We summarise the main characteristics of these simulations in Table~1 and refer the reader to \cite{2016MNRAS.457.4340K} for more details. 
We then select a subset of the available snapshots at different redshifts for the two simulations with the aim to reduce computational time. 
Eventually, we use 19 and 31 snapshots from BigMD27 and BigMDPL, respectively. The lower number of snapshots for BigMD27 is due to the smaller number of originally available snapshots.\footnote{The redshift snapshots for BigMD27 are: 0, 0.1399, 0.1759, 0.2702, 0.3581, 0.428, 0.4786, 0.4916, 0.5618, 0.5913, 0.6226, 0.6714, 0.7413, 0.8169, 0.8868, 1.445, 2.484, 2.891, 3.405. The redshift snapshots for BigMDPL are: 0, 0.03017, 0.04603, 0.08091, 0.1058, 0.1245, 0.1399, 0.1597, 0.1759, 0.1972, 0.2145, 0.237, 0.2702, 0.3153, 0.3581, 0.4037, 0.4401, 0.4786, 0.5191, 0.5618, 0.6069, 0.6548, 0.7053, 0.7593, 0.8169, 0.8369, 0.8868, 1, 1.445, 2.145, 2.484.} 
The number and distribution in redshift space is chosen requiring maximum deviations below few percents ($\lesssim 5$\%) at all redshifts in the cumulative mass function, and X-ray all-sky anisotropy angular power spectrum, with respect to what what would be obtained using all the available snapshots. To select this minimum acceptable number of snapshots, we analytically test the difference between i) the chosen subset, ii) all snapshots, and iii) the exact result obtained integrating a continuously evolving mass function in redshift space. The analytical mass function for this test is constructed using the on-line HMFcalc \citep{2013A&C.....3...23M} using the \cite{2010ApJ...724..878T} parameterisation. We compute the cumulative mass function, and the X-ray all-sky anisotropy angular power spectrum, using the phenomenological model of \cite{2013arXiv1311.4793Z,2015JCAP...09..060Z}, adopting different mass cuts and redshift bins.
Note that the difference with respect to the exact result, i.e., using a continuously evolving mass function in redshift space, can be larger than the above quoted 5\%. In fact, the limitation due to the original number of snapshots means that the resulting light-cones will be accurate, i.e, reproducing the exact result within few percents, only below $z \leq 0.94$ for BigMDPL and only within $ 0.16 \leq z \leq 0.86$ for BigMD27. Outside this range the difference can easily be higher than 10\% and may generate various artefacts depending on the mass and redshift range, and, if working in redshift bins, on the size of the redshift bins, e.g., the smaller the bins the higher the difference with respect to the exact solution, obviously. The systematic uncertainty in the cumulative mass function at low redshifts $z \leq 0.16$ for the WMAP-7 cosmology BigMD27 ranges from about 5\% to about 15\% for $M_{500}\geq5\times10^{13}$ and $M_{500}\geq5\times10^{14}$~$h_{0.7}^{-1}$~M$_\mathrm{sun}$, respectively. These numbers are to be kept in mind when looking at the comparison with the HIFLUGCS and REFLEX samples, which median redshifts are 0.05 and 0.08, respectively (see Section~\ref{sec:4}).

Taking as input the halo catalogues from the selected snapshots,\footnote{Note that the halos catalogues used for the BigMD27 and BigMDPL were obtained with the Bound Density Maxima \citep{1997astro.ph.12217K,2011arXiv1109.0003R} and Rockstar \citep{2013ApJ...762..109B} halo finders, respectively. While the latter catalogues contain the masses defined with respect to a density that is $\Delta=200$ and $500$ times the \emph{critical} density of the Universe, the first catalogues only contains $virial$ and $\Delta=200$-$critical$-masses, which we converted to $M_{500}$ using the \cite{2003ApJ...584..702H} method adopting the concentration-mass relation given by \cite{2011arXiv1104.5130P}. For simplicity, we do not implement the scatter in the \cite{2011arXiv1104.5130P} concentrations, which typically is, at a given redshift and mass, of about $0.12-0.15$ dex. We argue that this scatter has no significant impact on statistical studies in large mass bins as in the present paper except, perhaps, for some border effect at the lowest considered masses, where a cut is anyway artificially introduced to eliminate halos with $M_{500} < 10^{13}$~$h_{70}^{-1}$~M$_{\odot}$, and for the highest masses where the low numbers dominate the errors.} we then generate a hundred light-cones for each simulation by randomly changing the observer position in the simulation box. We follow the method of \cite{2010MNRAS.405..593Z} and \cite{2013MNRAS.429.1529F} in which the sky around the observer is divided in spherical shells corresponding to the redshift of each of the selected simulation snapshots. Each shell volume is then filled with non-overlapping copies of the corresponding simulation snapshot, randomly rotating and translating each copy. In doing so, the redshift of each halo in each snapshot is corrected considering its true line-of-sight distance with respect to the observer. Our 100-random-observer light-cones are, therefore, obtained through random rotations and translations of the simulation boxes.  A big difference with respect to \cite{2013MNRAS.429.1529F}, where the Millennium Simulation II with a volume of $(100~h^{-1}~\mathrm{Mpc})^{3}$ \citep{2009MNRAS.398.1150B} was used, is that here, thanks to the large volume of $(2.5~h^{-1}~\mathrm{Gpc})^{3}$ of the BigMD simulations, we do not need to replicate the simulation box but for the higher redshift ($z>1$) shells.

For both BigMD27 and BigMDPL, we chose among the 100 random observers the realisation closest to the analytical median of their cumulative mass function, checking several mass cuts. We do this by minimising the sum of the squared differences between the mass function in each redshift shell and the analytical median. The chosen realisation represents our baseline observer for which we will perform all calculations. For the other observers, as we will see, only a subset of quantities are computed in order to reduce computational time.

%%%%%%%%%%%%%%%%%%%%%%%%%%%%%%%%%%%%
%%%%%%%%%%%%%%%%%%%%%%%%%%%%%%%%%%%%
\subsection{Phenomenological modelling of gas density and temperature}
\label{sec:2.2}

The MultiDark N-body simulations contain information only on the dark matter properties of halos. Therefore, we need a model to assign to each halo in the light-cones a gas density and temperature. We describe in the following the steps of our phenomenological model built from state-of-the-art observations, which is inspired by the phenomenological approach of \cite{2013arXiv1311.4793Z}. Note that, for simplicity, we always assume spherical symmetry for the radial profiles.

\begin{enumerate}
\item Each halo in the light-cones is classified accordingly to four categories based on how disturbed its dark matter distribution is. The parameter used for this purpose is $X_\mathrm{off}$ which is the distance between the halo highest density peak and its centre of mass divided by its virial radius. We consider the distribution of $X_\mathrm{off}$ for halos with $M_{500} \geq 5 \times 10^{13}$~$h_{\mathrm{cosmo}}^{-1}$~M$_{\odot}$ in the snapshot at $z=0$, and define reference values by dividing it in quartiles. Thus, each subset consists of 25\% of the cluster population at $z=0$. In this way, we obtain four threshold values of $X_\mathrm{off}$ that we use across all redshifts in the light-cones to classify clusters as: very relaxed cool-cores, relaxed cool-cores, disturbed non-cool-cores, and very disturbed non-cool-cores. With this criterion, inspired by \cite{2010A&A...513A..37H} for the definition of four different temperature profiles of item~(iii) below, the fraction of disturbed clusters in our light-cones increases with redshift, and, at a fixed redshift, it increases at lower halo masses (note that we are talking about few percent differences at any fixed redshift snapshot with respect to the 25\% of the $z=0$ snapshot discussed above). We stress out that the choice of dividing the underlying cluster population at $z=0$ in quartiles is somewhat arbitrary and, while we believe is representative of the current theoretical and observational understanding (e.g.,~\citealp{2010A&A...513A..37H,2015ApJ...813L..17R,2017arXiv171205464S,2018MNRAS.474.1065S}), different scenarios could be investigated.\\
\item We use the two parameterisations taken from table~1 in \cite{2013A&A...550A.131P} for the profile of the pressure $P(r)$ for cool-core clusters and non cool-core clusters. This means that we use the same \emph{Planck} cool-core profile for both very relaxed and relaxed cool-cores, and the same \emph{Planck} non cool-core profile for both disturbed and very disturbed non-cool-cores. We will introduce further differences with the temperature profile in the next point. The mass and radius that go into the parameterisation refer to hydrostatic quantities. Therefore, we define $M_{500,\mathrm{HE}} = (1-b)~M_{500}$, where $M_{500}$ is the true mass of each halo as taken from the simulations halo catalogues, and we test different values for $1-b$, roughly spanning the parameter space considered in \cite{2016A&A...594A..24P}. We consider the values $0.6, 0.8, 1$ and $1.2$, and regard $1-b=0.8$ as our reference case. We note that the values $1-b=0.6$ and 1.2 are quite extreme, and that the first minimises the tensions between the \emph{Planck}-clusters-derived cosmological constraints and the analysis of the primary fluctuations in the cosmic microwave background (CMB). Note that in in our approach the $(1-b)$ factor encloses all possible deviations from the hydrostatic equilibrium as, e.g., the presence of non-negligible non-thermal pressure. We additionally introduce a Gaussian scatter, with $\sigma = 0.15$, in the natural logarithm of the normalisation of the pressure profile $P_{500}$, in order to reproduce the observed scatter around the $Y_\mathrm{SZ}-M_{500,\mathrm{HE}}$ scaling relation of \cite{2014A&A...571A..20P}, and we do include here the small deviation from self-similarity as given by equation~11 of \cite{2013A&A...550A.131P}.\\ 
\begin{figure}
\centering 
\includegraphics[width=.49\textwidth]{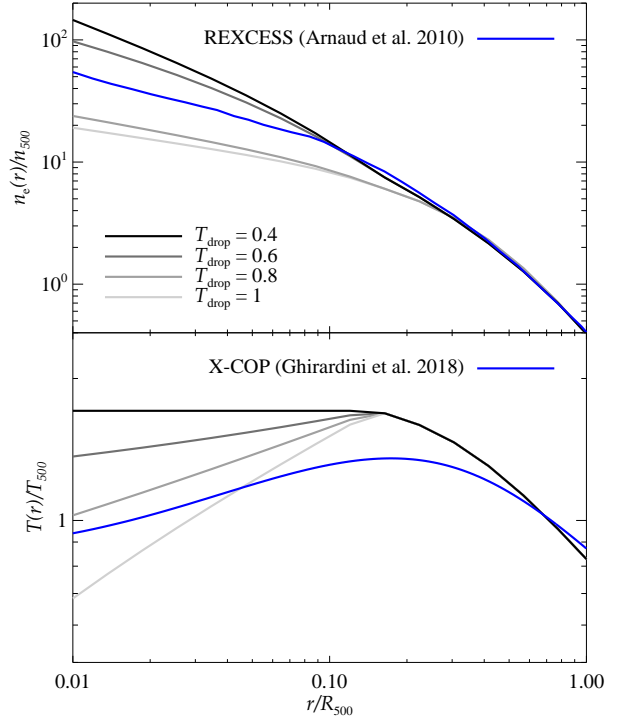}
\caption{\label{fig:gas_examples} Radial distribution of the ICM density (top) and temperature (bottom) profiles, divided by their volume-averaged values within $R_{500}$, for our definition of the dynamical status of clusters in the light-cones: very relaxed cool-cores $T_\mathrm{drop} = 0.4$, relaxed cool-cores $T_\mathrm{drop} = 0.6$, disturbed non-cool-cores $T_\mathrm{drop} = 0.8$, and very disturbed non-cool-cores $T_\mathrm{drop} = 1$. For comparison, we also show the mean density profile of the REXCESS sample \citep{2010A&A...517A..92A} and the temperature profile recently obtained from the X-COP sample \citep{2018arXiv180500042G}.} 
\end{figure}
\item As anticipated, the categories of clusters described above are further characterised by four different temperature profiles. In particular, we introduce a central temperature drop $T_\mathrm{drop} = T(r=0) / T(r=0.15 \times R_{500})$ inspired by \cite{2010A&A...513A..37H} with different values of $T_\mathrm{drop}$ for the four categories: very relaxed cool-cores $T_\mathrm{drop} = 0.4$, relaxed cool-cores $T_\mathrm{drop} = 0.6$, disturbed non-cool-cores $T_\mathrm{drop} = 0.8$ (practically a mild cool-core), very disturbed non-cool-cores $T_\mathrm{drop} = 1$. The temperature profile is constructed in few steps. The temperature profile is defined as:
\begin{eqnarray}
T(r\leq 0.01~R_{500}) & = & T_\mathrm{drop} \times T(r=0.15~R_{500}) \, ,\\
T(r\leq0.15~R_{500}) & = &  A+\left( B \times r^{0.24} \right) \, , \\
T(r\geq0.15~R_{500}) & = & T_\mathrm{norm} \times \left[ 1 + \left( \frac{r}{0.2~R_{200}} \right)^2 \right]^{-0.32}
\end{eqnarray}
where Eq.~(1) is kept fixed below $0.01 \times R_{500}$ just to avoid divergences. Eq.~(2) is the inner temperature decrease with power-law slope of 0.24 (adapted from the results of \citealp{2010A&A...513A..37H}), and parameters $A$ and $B$ are defined by smoothly connecting the three regimes in Eq.~(1), (2) and (3). Eq.~(3) accounts for the outer temperature profile decrease as in \cite{2010MNRAS.409..449P}. $T_\mathrm{norm} = \mathrm{norm} \times T_\mathrm{Mantz}$ is first set to the temperature obtained directly from the centrally-excised $M_{500}-T$ relation of \cite{2010MNRAS.406.1773M} or \cite{2016MNRAS.463.3582M} (i.e., $\mathrm{norm} = 1$), including the measured scatter and assuming that the Mantz et al.~masses are unbiased. Finally, we re-normalised the temperature such that: 
\begin{equation}
T_\mathrm{Mantz} = \frac{\int_{0.15~R_{500}}^{R_{500}} n_\mathrm{e}(r)^2~T(r)^{1/2}~T(r)~\mathrm{d}V}{\int_{0.15~R_{500}}^{R_{500}} n_\mathrm{e}(r)^2~T(r)^{1/2}~\mathrm{d}V} \, ,
\end{equation}
where $n_\mathrm{e}(r) = P(r)/kT(r)$ is the electron number density of the ICM. In Figure~\ref{fig:gas_examples}, we show the typical resultant gas density and temperature radial profiles volume-averaged within $R_{500}$.\footnote{With volume-averaged within $R_{500}$ we refer simply to the volume integral of the radial profile of a given quantity up to $R_{500}$ divided by the volume of a sphere of radius $R_{500}$.} We note that while our profiles are constructed starting from the \emph{Planck} pressure profiles and, therefore, from nearby objects \citep{2013A&A...550A.131P}, \cite{2018MNRAS.474.1065S} did not find significant deviations from the \cite{2010A&A...517A..92A} universal pressure profile from a higher redshift sample, and found good agreement with the electron density profiles of the REXCESS sample from \cite{2008A&A...487..431C}, from which the mean shown in blue in the top panel of Figure~\ref{fig:gas_examples} is constructed. A visual inspection of the temperature profiles in \cite{2018MNRAS.474.1065S} suggests also a good agreement with our temperature parametrisation choice. In the bottom panel of Figure~\ref{fig:gas_examples}, we eventually compare our temperature parametrisation with the profile recently obtained from the X-COP sample by \cite{2018arXiv180500042G}. Note the much shallower outer radial decline of the latter, also discussed by the authors in comparison to, e.g., \cite{2013SSRv..177..195R}, and the overall higher normalisation, of about 20\% at the peak value, of our temperature profiles. While the inner radial decline appears in agreement with our choices, it could be interesting to adapt our outer radial decline to the recent findings of \cite{2018arXiv180500042G} in our future releases.\\
\item We eventually calculate a number of quantities related to the X-ray and SZ properties of the clusters. We calculate the Xspec apec \citep{1996ASPC..101...17A} normalisation and volume-averaged temperature within $R_{500}$ that are used as inputs to calculate unabsorbed X-ray fluxes (in the observer-frame) and luminosities (in the rest-frame) within $R_{500}$. Note that, to speed-up calculations, we use the quantities integrated within $R_{500}$ as inputs for the apec model, rather than doing a radial integration of the apec output in several radial bins (the difference is typically below few percents). For simplicity, we always adopt a metallicity of 0.3 solar. Using the same inputs, we calculate the eROSITA count rate (in the observer-frame) in the $0.5-2$~keV energy range within $R_{500}$ following the procedure of \cite{2012MNRAS.422...44P}, in particular assuming for simplicity an uniform sky coverage and hydrogen column density of $3 \times 10^{20}$~atom~cm$^{-2}$ for the photoelectric absorption, but using the latest eROSITA response functions for the sensitivity and effective area.\footnote{https:/wiki.mpe.mpg.de/eRosita/erocalib$\_$calibration}
Note that we do not include the expected instrumental background in our  eROSITA count-rate simulations.
We calculate the $Y_\mathrm{SZ}$ signal (as in equation~5 of \citealp{2014A&A...571A..20P}) both within $R_{500}$ and within $R_{500,\mathrm{HE}}$, the latter for a direct comparison with the \emph{Planck} results. Note that the difference between the two integration limits is about 6\% for $1-b =0.8$. We eventually calculate $M_\mathrm{gas,HE}$, $f_\mathrm{gas}=M_\mathrm{gas,HE}/M_{500, \mathrm{HE}}$, and $Y_\mathrm{X} = M_\mathrm{gas,HE} \times T_{500, \mathrm{HE}}$ using as reference integration radius $R_{500, \mathrm{HE}}$ for comparison with the \cite{2009ApJ...692.1033V} results.\footnote{Note that \cite{2009ApJ...692.1033V} derives the temperature as the average projected temperature within $0.15 \times R_{500} - R_{500}$. In our case, the difference between the latter and the volume-averaged $T_{500}$ is below 1\%.}\\
\item As part of this project, we are interested in looking at the angular power spectra of the X-ray luminosity and SZ signal, as well as their cross-correlation. Therefore, we generate the 0.5-2~keV and SZ profiles of our clusters projected onto the sky. We calculate the corresponding surface brightness in 10 spherical shells at $r_{i=0-9} = (\Delta r \times i) + \Delta r/2$ with $\Delta r = R_{500}/10$. We then calculate the corresponding projected X-ray flux and SZ signal in each shell as:
\begin{eqnarray}
F_{0.5-2,i}(\tilde{r}) & = & 2~\pi~r_i~\Delta r \int_{r_i}^{R_{500}} \frac{2~F_{0.5-2,i}(\tilde{r})~\tilde{r}}{ \sqrt{\tilde{r}^{2}-r_i^2}} \mathrm{d}\tilde{r} \, , \\
Y_{\mathrm{SZ},i}(\tilde{r}) & = &  2~\pi~r_i~\Delta r \int_{r_i}^{R_{500}} \frac{2~Y_\mathrm{SZ,i}(\tilde{r})~\tilde{r}}{ \sqrt{\tilde{r}^{2}-r_i^2}} \mathrm{d}\tilde{r} \, ,
\end{eqnarray}
where $F_{0.5-2,i}(\tilde{r})$ and $Y_\mathrm{SZ,i}(\tilde{r})$ are the radial profiles. For the case of the X-ray flux, we calculate the apec normalisation and temperature inputs for the volume corresponding to each one of the 10 spherical shells, and calculate with Xspec the corresponding $F_{0.5-2,i}$. If Eq.~(5) and (6) are summed over all $i$ one gets back the total $F_{0.5-2}$ and $Y_\mathrm{SZ}$, respectively, volume-integrated within $R_{500}$. The fact that the projected profiles up to $R_{500}$ are normalised to give back the volume-integrated quantities within $R_{500}$ implies that we have implicitly assumed that cluster quantities do not extend beyond $R_{500}$. This is a practical choice made to speed-up calculations that we will, however, relax in some of the cases explored in Section~\ref{sec:5}.
To obtain the projected profile of the observed counts, used in Section~\ref{sec:5}, we just re-normalise $F_{0.5-2,i}$ to the volume-integrated count rate within $R_{500}$.\\
\item Throughout this paper, two different versions of the light-cone catalogues are used (these different versions are also available on-line; see Appendix~A). For the first set of catalogues that include only clusters with $M_{500}\geq5\times10^{13}$~$h_\mathrm{cosmo}^{-1}$~M$_\mathrm{sun}$, we follow the procedure described above. We then use a second set of catalogues that include clusters down to $M_{500} = 1\times10^{13}$~$h_\mathrm{cosmo}^{-1}$~M$_\mathrm{sun}$ and are, therefore, significantly heavier to handle, both in terms of storage and computational time. In this case, we follow a simplified procedure after having performed step (i) in the same way as above. We use precomputed tables providing $Y_\mathrm{SZ}$, X-ray fluxes, X-ray luminosities, count rates, and projected-X-ray and SZ profiles. Each tabulated quantity is computed in terms of $z$, $M_{500}$, and $T_\mathrm{drop}$. Therefore, by interpolation, it is possible to compute the properties of each cluster according to its redshift, mass and $T_\mathrm{drop}$. The interpolation tables are calculated from quantities without scatter on $P_{500}$ and in the $M_{500}-T$ relation, and the deviation from the full calculation is below 1\%. The scatter is added to $Y_\mathrm{SZ}$ and the X-ray luminosities. The X-ray fluxes, count rates, and projected profiles are eventually re-normalised to the quantities that include the scatter. We introduce a scatter of $0.065$~dex on $Y_\mathrm{SZ}$, and of $0.18$~dex on the X-ray luminosities where the four different cluster populations give a final scatter around $0.2$~dex (see Section~\ref{sec:3}). The scatter values are kept fixed at all redshifts and are independently added to the SZ signal and X-ray luminosities, i.e., two different random Gaussian numbers are used.\\ 
\item We adopt the simplified procedure of step (vi) also for the 100 random observers mentioned in Section~\ref{sec:2.1} that we use to partially evaluate the cosmic variance due to other observers in the simulation box. However, in this case, we limit the number of calculated quantities both for speeding-up calculations and for economy of disk space. In particular, we do not calculate the projected X-ray and SZ profiles. When needing these, e.g., in calculating the X-ray angular power spectrum in Section~\ref{sec:5}, we define four median projected profiles from the ones calculated for the baseline-observer light-cone for each cluster type (one for each $T_\mathrm{drop}$), and then scale them using the total fluxes of each cluster in the 100-observers light-cones. The deviations of the so-obtained X-ray projected profiles with respect to the full calculation or the simplified approach of point (vi) is not particularly severe, but tend to be larger for larger radii and smaller masses connected to the role of a lower temperature in the strength of emission lines. A full test using both approaches with the same catalogue showed differences of the median-derived X-ray projected profiles with respect to the correct ones, for all objects with $M_{500}\geq10^{13}$~$h_\mathrm{cosmo}^{-1}$~M$_\mathrm{sun}$, at the level of 3\%, 10\% and 30\% (this is the 90\%-quantile of the distribution of deviations) at $0.1$, $0.5$ and $1\times R_{500}$, respectively. While we did not use the SZ projected profile in this work and, therefore, did not perform the same check here, we would expect the deviations in that case to be much smaller due to the more regular dependency on the temperature. At any rate, for the scope of this paper that is to evaluate the X-ray angular power spectrum of the eROSITA-detected and undetected clusters, we found no significant difference between the two approaches in the $C_l$-plots shown in Section~\ref{sec:5}.
\end{enumerate}

Note that, in constructing the pressure, gas and temperature profiles, we self-consistently take into consideration the underlying cosmology parameters of each BigMD simulation by adopting the proper values in the redshift evolution parameter $E(z)^{2}=\Omega_\mathrm{M}(1+z)^{3}+\Omega_\Lambda$, and by correcting, where necessary, the value of the adopted Hubble constant. However, in the following section, we show figures where the Hubble constant explicitly present in the units is always $0.7$ for comparison with existing literature. Note that even considering the different cosmologies, there will be no one-to-one correspondence between BigMD27 and BigMDPL. The differences are very small and are due to the fact that the cluster population, and their separation in the four dynamical states that we defined, is different in the two different simulations and their light-cones. Additionally, the difference in the X-ray luminosity is more pronounced as the apec model requires the distance as input, and its cosmology dependence will remain encoded in the outputs (fluxes, luminosities, counts) despite final corrections on $E(z)$ and/or $h_\mathrm{cosmo}$ for plotting purposes.

%%%%%%%%%%%%%%%%%%%%%%%%%%%%%%%%%%%%
%%%%%%%%%%%%%%%%%%%%%%%%%%%%%%%%%%%%
\section{Validation of the Scaling Relations for the Light-cones}
\label{sec:3}

The phenomenological modelling described in the previous section has been constructed with the objective of reproducing current X-ray and SZ observations. In particular, it should reproduce a set of state-of-the-art scaling relations for well known galaxy cluster samples, in order to have reliable predictions once extended into the unknown, given certain assumptions as a self-similar redshift evolution. For the following validation of the scaling relations we use the light-cones containing clusters with $M_{500}\geq5\times10^{13}$~$h_\mathrm{cosmo}^{-1}$~M$_\mathrm{sun}$ obtained with the full calculation of points (i)-(v) of Section~\ref{sec:2.2}. We compare the results from our light-cones for $0.1 \leq z \leq 0.3$ and $M_{500}\geq10^{14}$~$h_{70}^{-1}$~M$_\mathrm{sun}$ to our reference observational scaling relations from \cite{2009ApJ...692.1033V}, \cite{2014MNRAS.438...78R}, \cite{2016MNRAS.463.3582M} and \cite{2017MNRAS.469.3738S}. In particular, we choose \cite{2014MNRAS.438...78R} as they constructed a set of scaling relations of different quantities (X-ray luminosity, $Y_\mathrm{X}$ and $Y_\mathrm{SZ}$-to-mass) that are self-consistent. The redshifts and masses have been chosen to roughly match the currently most used X-ray observational samples of clusters. The light-cone scaling relations shown in the figures and discussed in the following are obtained by a linear fit to all the clusters in the chosen sub-sample.

\begin{figure}
\centering
\includegraphics[width=.49\textwidth]{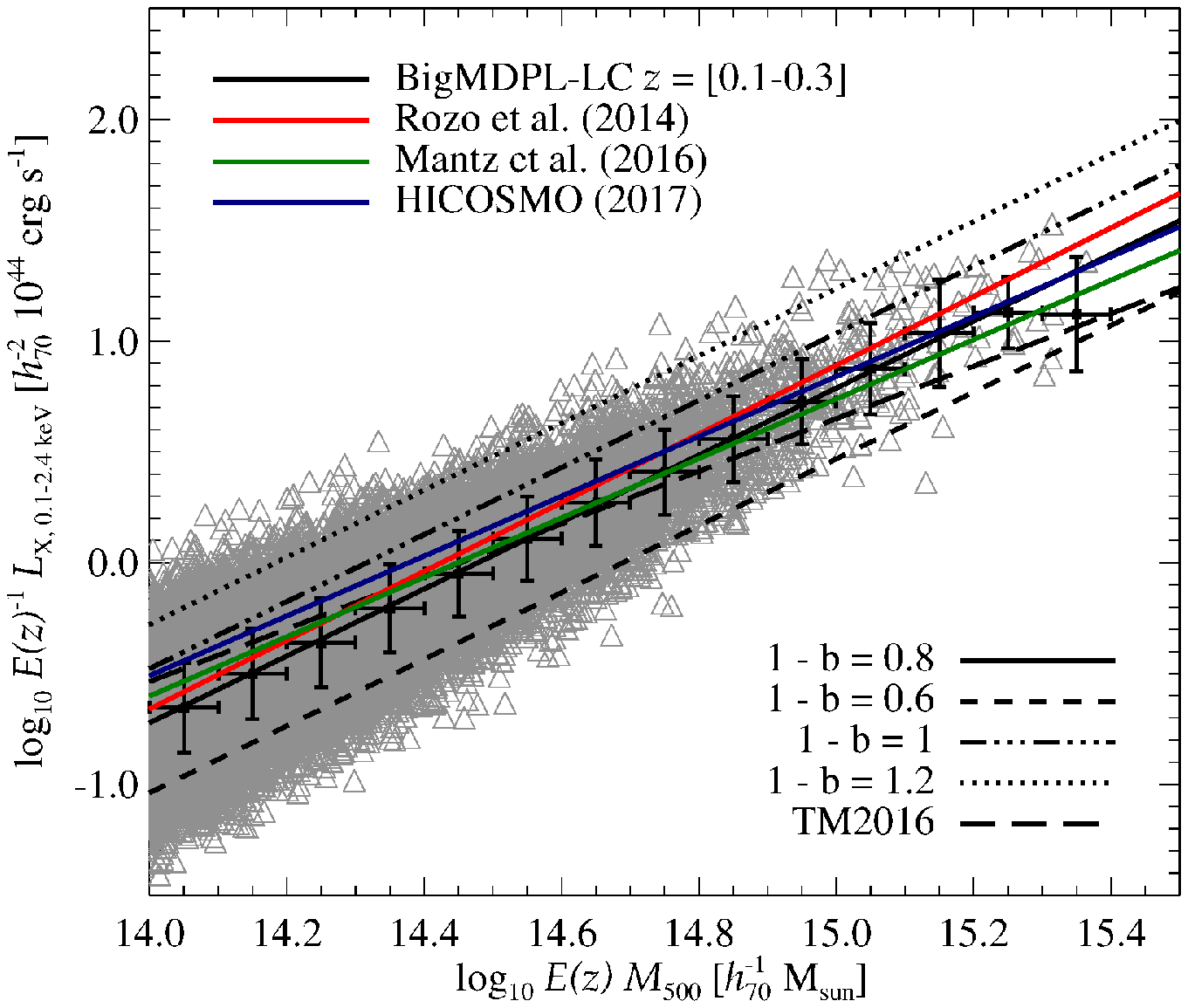}
\caption{\label{fig:LM} X-ray luminosity, in the $0.1-2.4$~keV energy range, -to-mass scaling relations for our BigMDPL light-cones against the observational scaling relations from \protect\cite{2014MNRAS.438...78R}, \protect\cite{2016MNRAS.463.3582M} and \protect\cite{2017MNRAS.469.3738S}, the last indicated as HICOSMO in the legend (we show their bias-corrected scaling relation for the whole sample). The grey triangles are the galaxy clusters in the BigMDPL-LC with $1-b=0.8$ adopting the $M_{500}-T$ relation of \protect\cite{2010MNRAS.406.1773M} where the black crosses represent the medians and root mean square (rms) in different mass bins. The scaling relations are obtained via linear fitting of all clusters in the chosen redshift and mass sub-sample (i.e., the triangles). The acronym TM2016 in the bottom-right legend refers to the light-cones with $1-b =0.8$ adopting the \protect\cite{2016MNRAS.463.3582M} $M_{500}-T$ relation.}
\end{figure}

Figure~\ref{fig:LM} shows the scaling relation of the X-ray luminosity, in the $0.1-2.4$~keV energy range, as a function of $M_{500}$. We show the results for the light-cones (LC) constructed from BigMDPL with $1-b = 0.8$, 0.6, 1, 1.2 adopting the $M_{500}-T$ relation of \cite{2010MNRAS.406.1773M}, and with $1-b =0.8$ and adopting the \cite{2016MNRAS.463.3582M} $M_{500}-T$ relation. The results obtained with the light-cones constructed from BigMD27-LC are almost the same and the resulting scaling relations are shifted down by less then 0.04~dex. The models that best compare with the \cite{2014MNRAS.438...78R}, \cite{2016MNRAS.463.3582M} and \cite{2017MNRAS.469.3738S} observational scaling relations are our baseline model, i.e., the one with $1-b =0.8$ and the \cite{2010MNRAS.406.1773M} $M_{500}-T$, and the one with the most recent \cite{2016MNRAS.463.3582M} $M_{500}-T$ relation. The models with $1-b = 0.6$, $1$ and $= 1.2$ have the effect of scaling down/up the X-ray luminosity at parity of $M_{500}$ with respect to the case with $1-b =0.8$ and, therefore, are significantly off with respect to our reference observations. The scatter around the scaling relations is about $0.2$ dex in luminosity in all cases, which compares well with observations, while being a bit larger than typical values \citep{2010MNRAS.406.1773M,2014MNRAS.438...62R} due to the combined effect of the scatter on $P_{500}$ and in the temperature-mass relation.

\begin{figure*}
\centering
\includegraphics[width=.49\textwidth]{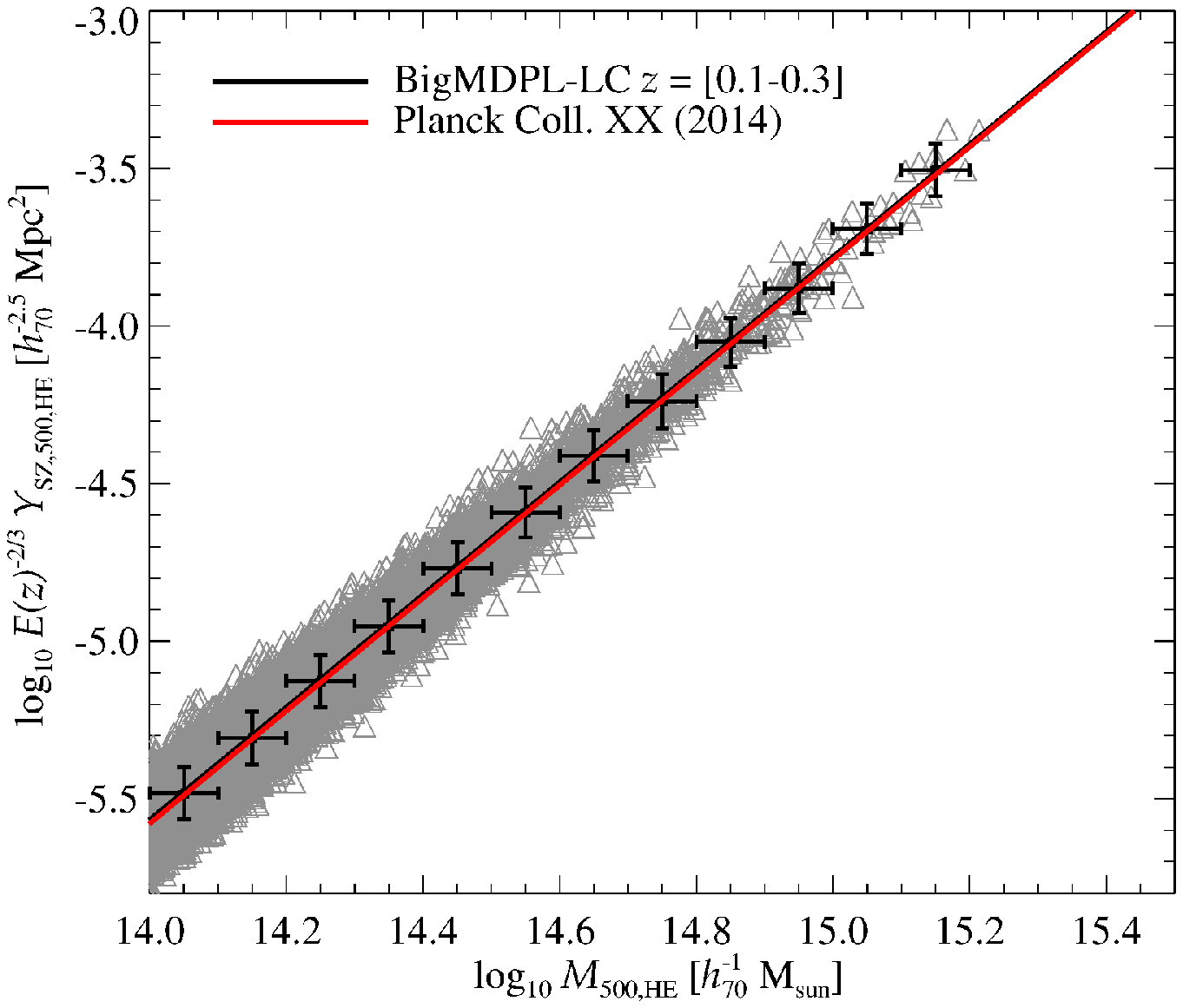}
\includegraphics[width=.49\textwidth]{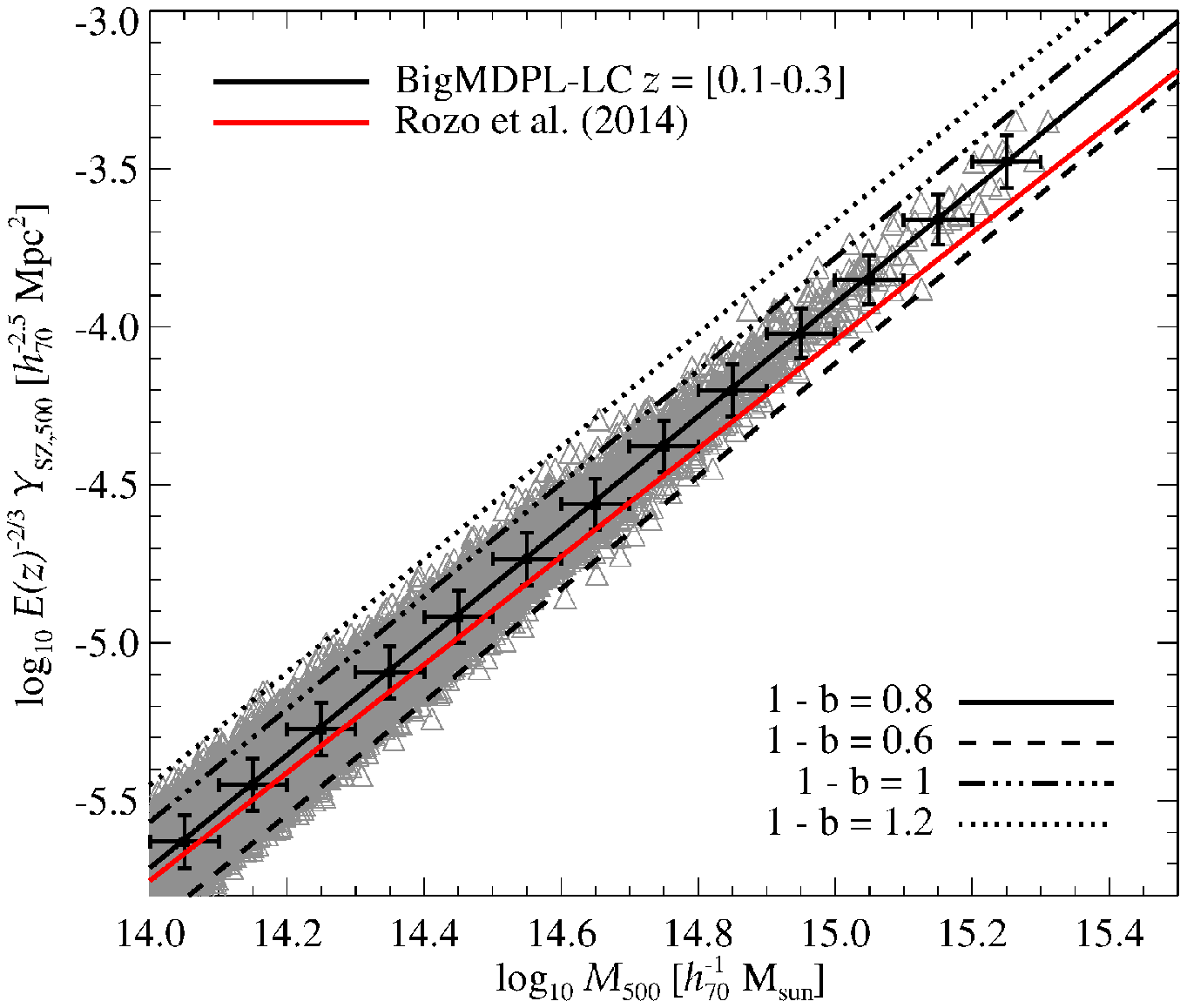}
\caption{\label{fig:SZ} $Y_\mathrm{SZ}$-to-mass scaling relations for our BigMDPL light-cones against observational scaling relations from \protect\cite{2014A&A...571A..20P} and \protect\cite{2014MNRAS.438...78R}. The grey triangles are the galaxy clusters in the BigMDPL-LC with $1-b=0.8$ where the black crosses represent the medians and rms in different mass bins. The scaling relations are obtained via linear fitting of all clusters in the chosen redshift and mass sub-sample (i.e., the triangles). The left panel shows the $Y_\mathrm{SZ}$--$M_{500,\mathrm{HE}}$ scaling relations, while the right panel shows the $Y_\mathrm{SZ}$--$M_{500}$ one. Therefore, in the left panel there is only one line for BigMDPL-LC, while in the right panel there is no line for TM2016 as in Figure~\ref{fig:LM} because the  choice of temperature-mass relation has no impact here.}
\end{figure*}

Figure~\ref{fig:SZ} shows the $Y_\mathrm{SZ}$-to-mass scaling relations for the clusters in BigMDPL-LC (the BigMD27-LC results are basically the same and shifted down by about 0.02~dex). There are two panels that show the scaling relation with respect to $M_{500,\mathrm{HE}}$, and the corresponding comparison with \cite{2014A&A...571A..20P} (left panel), and with respect to $M_{500}$ comparing with \cite{2014MNRAS.438...78R} (right panel). We almost perfectly match the \cite{2014A&A...571A..20P} results which should come as no surprise give that we built our models using the \emph{Planck} pressure profiles \citep{2013A&A...550A.131P}. We compare well also with the \cite{2014MNRAS.438...78R} scaling relation when using the true (i.e., unbiased) mass of the clusters with our baseline ($1-b =0.8$) model, while models with different values of the hydrostatic mass bias are significantly off the observational scaling relation. The scatter around the scaling relations is $0.065$ dex in all cases, matching well the observations.

\begin{figure*}
\centering
\includegraphics[width=.49\textwidth]{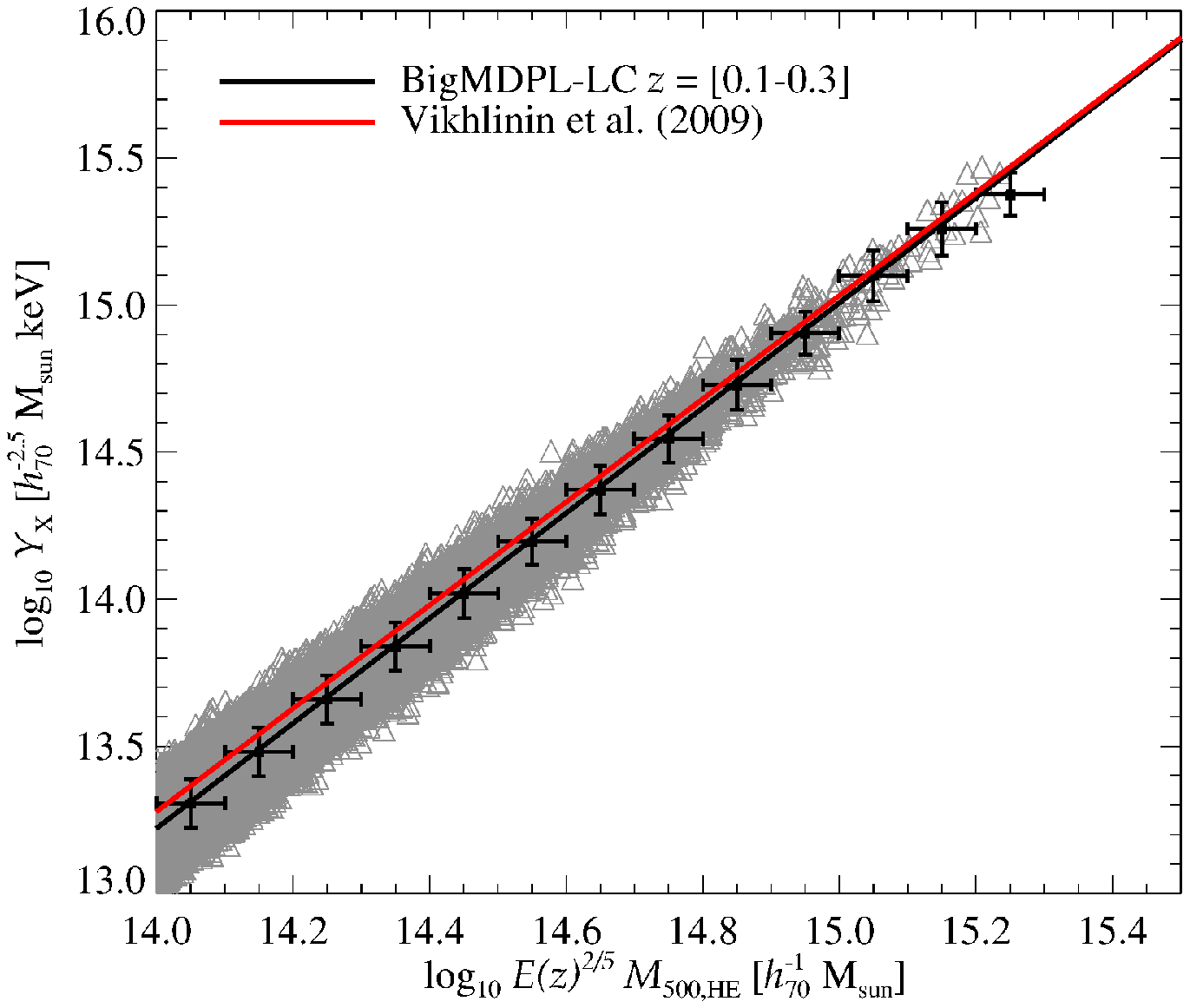}
\includegraphics[width=.49\textwidth]{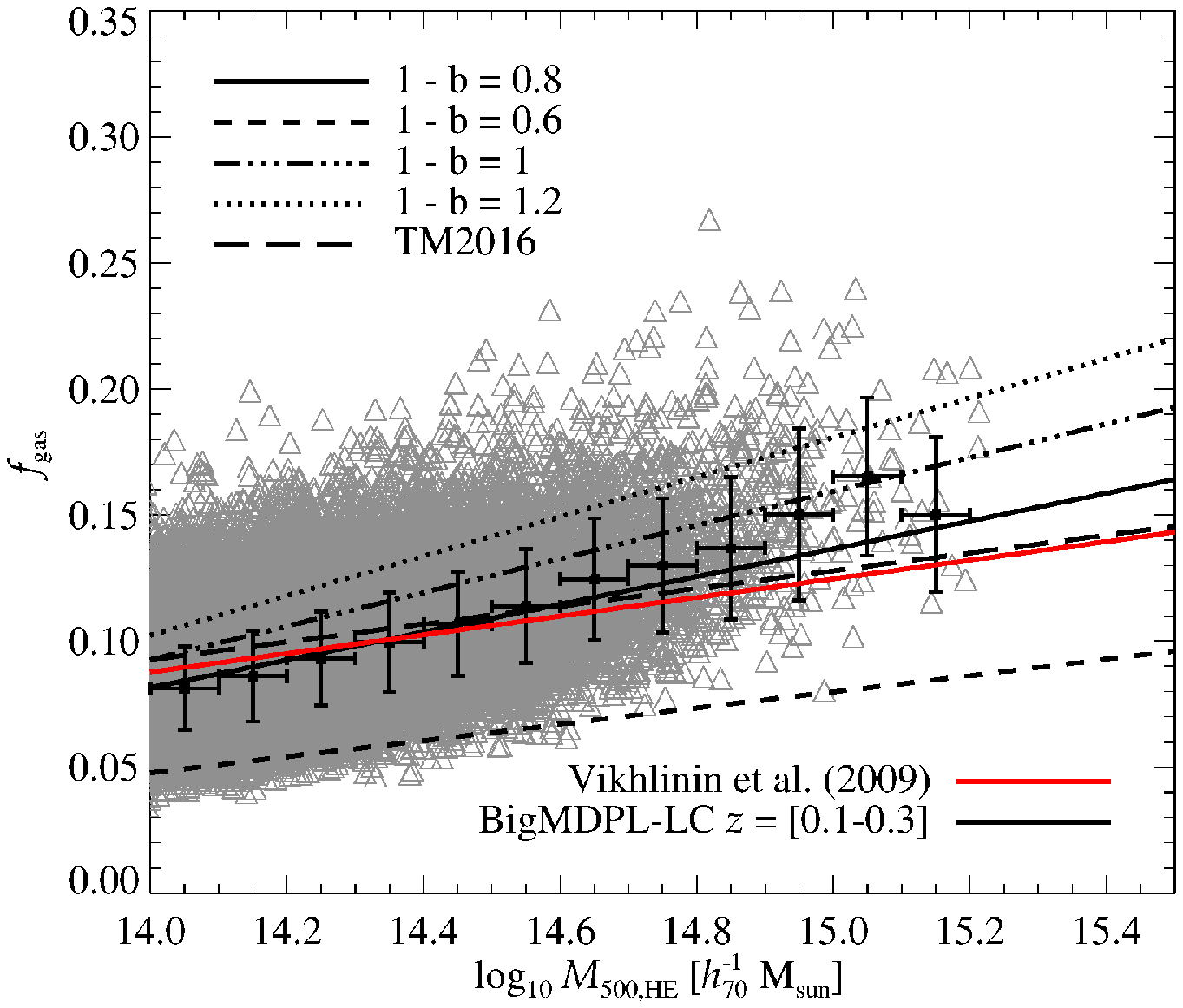}
\caption{\label{fig:yx_gas} $Y_\mathrm{X}$ and $f_\mathrm{gas}$-to-$M_{500,\mathrm{HE}}$ scaling relations for our BigMDPL light-cones against observational scaling relations from \protect\cite{2009ApJ...692.1033V}. The grey triangles are the galaxy clusters in the BigMDPL-LC with $1-b=0.8$ where the black crosses represent the medians and rms in different mass bins. The scaling relations are obtained via linear fitting of all clusters in the chosen redshift and mass sub-sample (i.e., the triangles). In the left panel, showing $Y_\mathrm{X}$--$M_{500,\mathrm{HE}}$, there is only one line for BigMDPL-LC because when calculating $Y_\mathrm{X} = M_\mathrm{gas,HE} \times T_{500, \mathrm{HE}}$ the dependences on the hydrostatic bias ($1-b$) and temperature cancel out. However, in the right panel, showing $f_\mathrm{gas}$--$M_{500,\mathrm{HE}}$, the five models are separated as those dependencies remains encoded in $f_\mathrm{gas}=M_\mathrm{gas,HE}/M_{500, \mathrm{HE}}$.}
\end{figure*}

Figure~\ref{fig:yx_gas} shows the scaling relations for $Y_\mathrm{X}$ and $f_\mathrm{gas}$-to-$M_{500,\mathrm{HE}}$ for the BigMDPL-LC (the BigMD27-LC results are basically the same and shifted down by about 0.02~dex in both cases) compared with the \cite{2009ApJ...692.1033V} results. Both scaling relations from our light-cones compare well with observations. 
We note that recently \cite{2017arXiv171105344D}, using a sample of weak-lensing mass calibrated SPT clusters, found good agreement with the \cite{2009ApJ...692.1033V} results, while showing a somewhat steeper $M_\mathrm{gas}$-to-mass relation, similarly to what was found by \cite{2017arXiv171100917C} for the $f_\mathrm{gas}$-to-mass relation.
At any rate, the scatter around the $Y_\mathrm{X}$--$M_{500,\mathrm{HE}}$ relation is 0.065~dex, somewhat larger than $\le 0.05$~dex found by \cite{2009ApJ...692.1033V} and others \citep{2014MNRAS.438...62R} due to the scatter that we introduced on $P_{500}$ in order to reproduce the $Y_\mathrm{SZ}$--$M_{500,\mathrm{HE}}$ scatter, propagating into the calculation of $M_\mathrm{gas,HE}$. The right panel of Figure~\ref{fig:yx_gas} shows five well separated scaling relations for our five models despite the fact that we are plotting $f_\mathrm{gas}$ against $M_{500,\mathrm{HE}}$ and, therefore, the hydrostatic bias and temperature should not affect this. However, we calculate $M_\mathrm{gas,HE}$ from the ICM density that we obtain from the pressure profile dividing by the temperature profile, $n_\mathrm{e}(r) = P(r)/kT(r)$, and, therefore, the hydrostatic bias and temperature dependencies remain encoded in $f_\mathrm{gas}=M_\mathrm{gas,HE}/M_{500, \mathrm{HE}}$. Another notable feature of the $f_\mathrm{gas}$-to-mass relation is that the distribution of clusters around the scaling relation is skewed toward higher values of $f_\mathrm{gas}$, and this feature becomes more noticeable as the mass of clusters increases. We interpret this as due the fraction of disturbed clusters that increases at higher masses and redshifts with respect to the 50\% fraction fixed at the $z=0$ snapshot (see Section~\ref{sec:2.2}). 

A final note on the redshift evolution. The light-cones are constructed such that all observables evolve self-similarly, with the exception of the small deviation from self-similarity introduced on $P_{500}$ following \cite{2013A&A...550A.131P}. Therefore, the X-ray luminosity $L_{\mathrm{X,0.1-2.4~keV}}$ evolves as $E(z)^{7/3}$, $Y_\mathrm{SZ}$ and $Y_\mathrm{X}$ evolve as $E(z)^{2/3}$ and $E(z)^{2/5}$, respectively, while $f_\mathrm{gas}$ does not show any redshift evolution.

\begin{table*}
\begin{center}
\caption{\label{tab:halo_numbers} Number of detected eROSITA galaxy clusters in the different BigMD light-cones.}
\resizebox{\textwidth}{!}{
\begin{tabular}{lcccccccc}
\hline\hline
\phantom{\Big|}
             & BigMDPL-LC  &                   &               &                  &               & BigMD27-LC  & BigMDPL-LC            \\
Sample &  $1-b=0.8$      & $=0.6$ & $=1$ & $=1.2$ & TM2016 & $1-b=0.8$       & $1-b=0.8$ 100 observers \\
\hline\\[-0.5em]
HIFLUGCS (64 clusters)  &  141  &  59   &  296   &  567   &  195  &  109  &  134$\pm$15 (11\%)  \\
REFLEX (452 clusters)  & 861 &  325  &  1788   &   3229   &  1239   &  632  &  855$\pm$26 (3\%) \\
\hline\\[-0.5em]
eROSITA $M_{500}\geq1\times10^{13}$     &  183555     &    78155    &     347298    &   575201   &  355611    &  135132    & 183793$\pm$761 (0.4\%) \\
eROSITA $M_{500}\geq5\times 10^{13}$    &   128623     &    58894   &    225675     &  344082     &   188168   &  95882    &  128898$\pm$432 (0.3\%)\\
eROSITA $M_{500}\geq1\times 10^{14}$    &    87446    &   43815    &   137716      &   185336    &  101453    &  66017    &  87540$\pm$298 (0.3\%) \\
eROSITA $M_{500}\geq5\times 10^{14}$    &    2622    &   2580    &    2626    &   2626    &  2619     & 2388     &  2606$\pm$42 (1.6\%) \\
\hline
\end{tabular}
} %end of resize box
\end{center}
\small{{\bf Note.} The top two rows show the number of detected objects when applying the selection criteria of the HIFLUGCS ($F_{0.1-2.4~\mathrm{keV}} \geq 2\times10^{-11}$~erg~cm~$^{-2}$~s$^{-1}$, 65\% of the sky with 100\% completeness; \citealp{2002ApJ...567..716R}) and REFLEX ($F_{0.1-2.4~\mathrm{keV}} \geq 3\times10^{-12}$~erg~cm~$^{-2}$~s$^{-1}$, 34\% of the sky with 90\% completeness; \citealp{2001A&A...369..826B}) samples, where the numbers between parenthesis in the first column are the numbers obtained in the observational samples. The bottom rows show the number of clusters detected by eROSITA defined by adopting the criterium of having $\geq 50$ counts in the $0.5-2$~keV energy range within $R_{500}$ after 1.6~ks of observations and considering only 0.658 of the sky, i.e., excising $\pm20^\circ$ from the Galactic Plane. The different rows refer to different low mass cuts in units of $h_\mathrm{cosmo}^{-1}$~M$_\mathrm{sun}$. The last column shows the results for the 100 random observers quoting the average (we find no significant difference with respect to calculating the median) and RMS, also quoted in percentage in parenthesis.} 
\end{table*}

%%%%%%%%%%%%%%%%%%%%%%%%%%%%%%%%%%%%%%%%%%%%%%%%%%%%%%%%%%%%%%%%%%%
%%%%%%%%%%%%%%%%%%%%%%%%%%%%%%%%%%%%%%%%%%%%%%%%%%%%%%%%%%%%%%%%%%%
\section{The eROSITA all-sky survey}
\label{sec:4}

\begin{figure*}
\centering
\includegraphics[width=.49\textwidth]{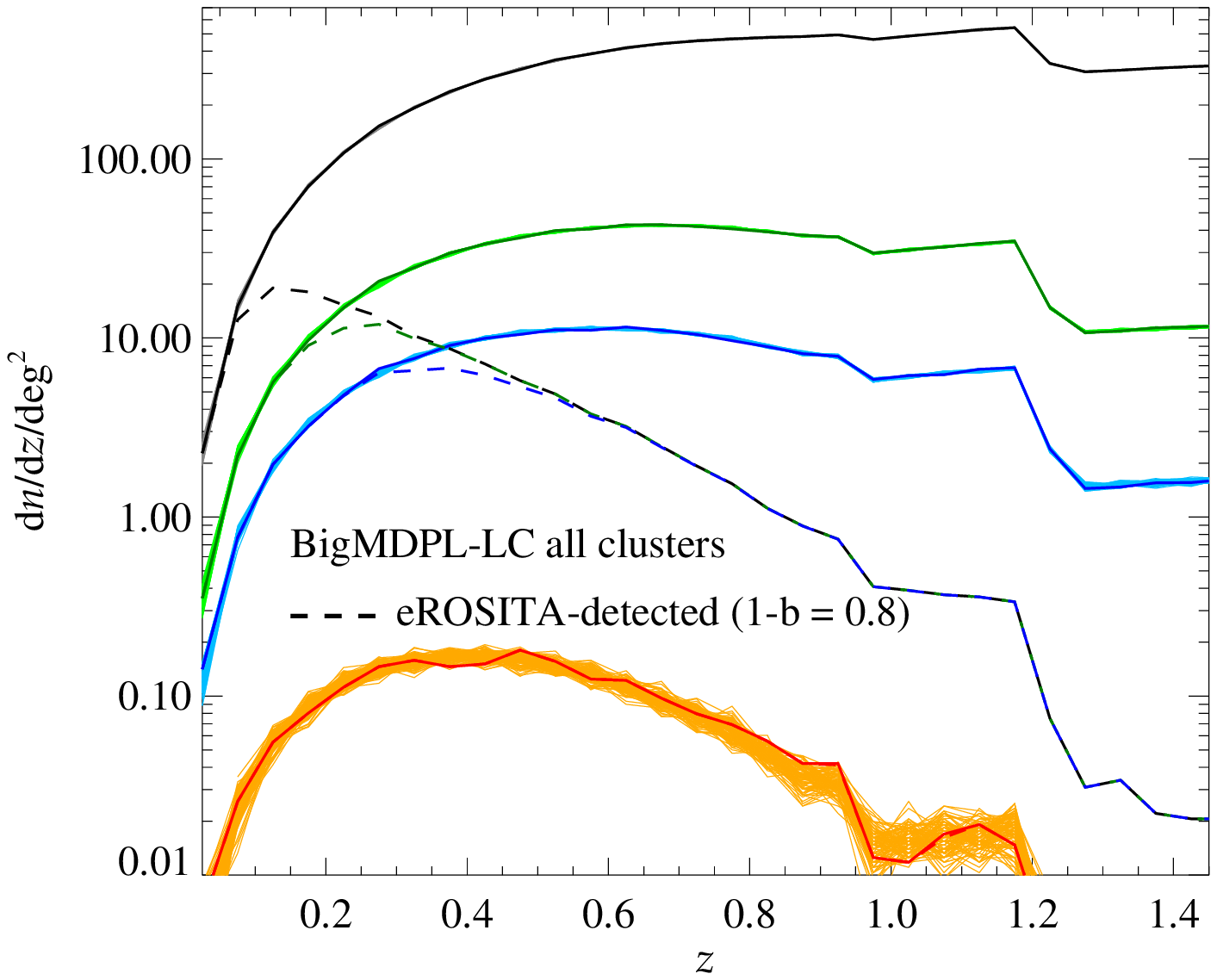}
\includegraphics[width=.49\textwidth]{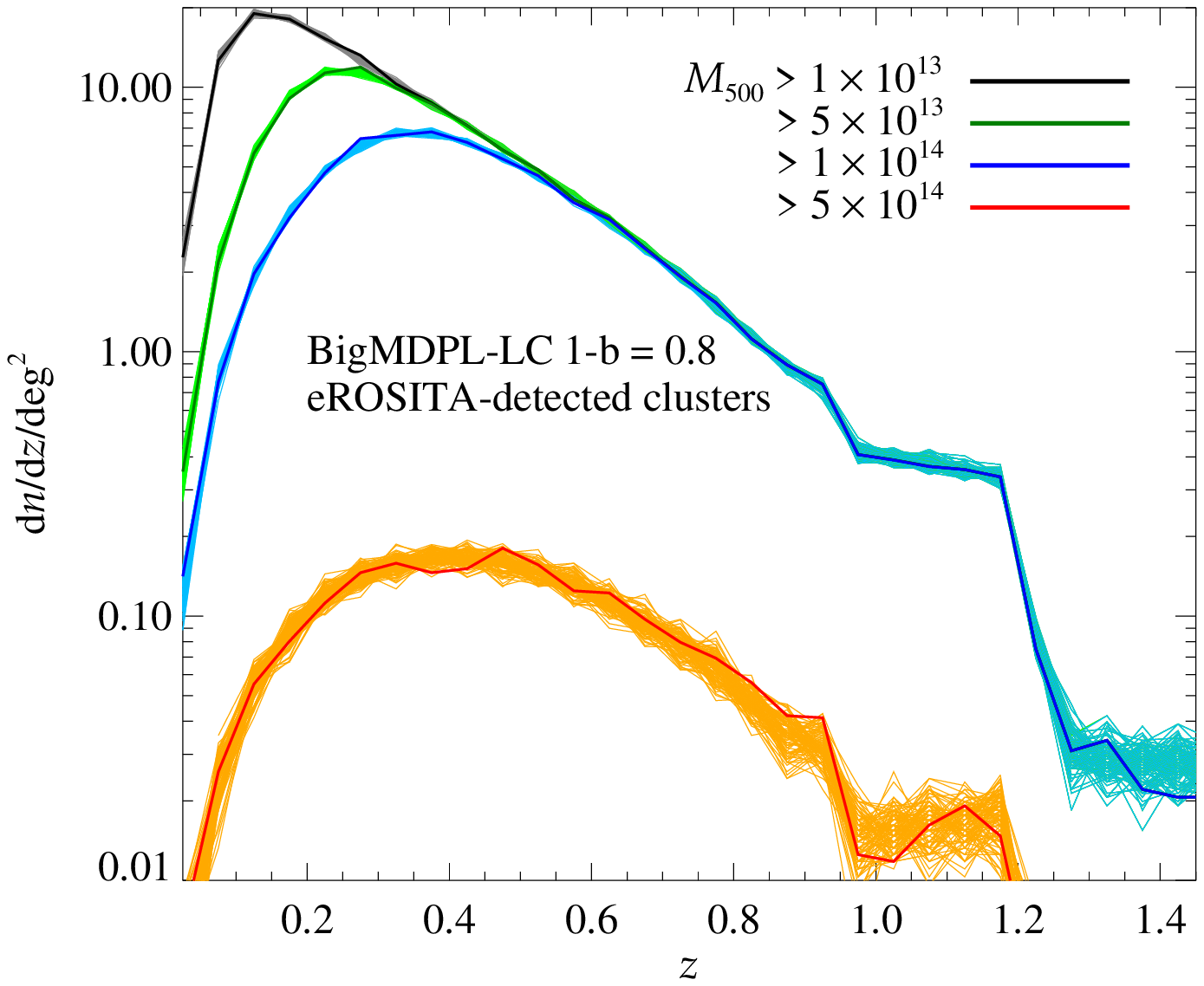}
\caption{\label{fig:num_halos_LC_1} The left panel shows the total number of clusters per redshift bin ($\Delta z = 0.05$) and square degree in our baseline BigMDPL light-cone (thick lines), together with the variation among the 100 random observers, and the corresponding number of clusters detected by eROSITA for different mass cuts in units of $h_\mathrm{cosmo}^{-1}$~M$_\mathrm{sun}$ (colour-coded as in the right panel). The right panel shows the zoomed-in version of the number of clusters detected by eROSITA with their variation among the 100 random observers. Note in both panels the artefacts appearing at $z \gtrsim 0.94$ due to the limited number of available redshift snapshots in the original BigMDPL simulation at high redshifts (see Section~\ref{sec:2.1}).}
\end{figure*}

In this section we present some predictions for the number of clusters detectable by the eROSITA all-sky X-ray survey following closely the approach of \cite{2012MNRAS.422...44P}. In particular, as already mentioned in Section~\ref{sec:2.2}, in calculating eROSITA counts we assume, for simplicity, an uniform sky coverage and hydrogen column density for the photoelectric absorption. We defined as detected all clusters with counts $\geq 50$, in the $0.5-2$~keV energy range within $R_{500}$, after 1.6~ks of eROSITA observations. 
We are aware that this is a simplified criterion, however, recent detailed simulations of the eROSITA extra-galactic sky by \cite{2018arXiv180608652C} shown that it is indeed a good approximation of the eROSITA selection function that suffices the scopes of the present paper. Additionally, \cite{2018arXiv180608652C} found results similar to \cite{2012MNRAS.422...44P}, which also adopted a simple 50-counts limit, and therefore, further motivate our choice, also for consistency with previous work.
Note that in calculating the eROSITA counts for our light-cones we did not include Poissonian noise, and the power spectrum study of count-rate sky maps of Section~\ref{sec:5} does not include it either. The reason for this choice is to be able, if needed, to generate a posteriori Poissonian realisations of the results.

Table~\ref{tab:halo_numbers} contains the number of clusters detected by eROSITA for some of our BigMD light-cones. We show all models with the \emph{Planck}-cosmology light-cone BigMDPL-LC together with the baseline model (with $1-b=0.8$) for the WMAP-cosmology light-cone BigMD27-LC. In doing so we use as starting point the light-cones including all halos with  $M_{500}\geq 10^{13}$~$h_\mathrm{cosmo}^{-1}$~M$_\mathrm{sun}$ (see point (vi) of Section~\ref{sec:2.2}). We also show the variation in such numbers among the 100 random observers for the BigMDPL-LC. We complement this with a comparison with two well-known clusters samples, HIFLUGCS \citep{2002ApJ...567..716R} and REFLEX \citep{2001A&A...369..826B}, adopting their sky coverage and reported completeness above the $0.1-2.4$~keV flux cuts, $2\times10^{-11}$ and $3\times10^{-12}$~erg~cm~$^{-2}$~s$^{-1}$, respectively. We remind the reader that the median redshift for the HIFLUGCS and REFLEX samples is 0.05 and 0.08, respectively.

Most of our light-cones systematically over-predict the counts statistics obtained by HIFLUGCS and REFLEX, also when considering the root-mean-square (RMS) variation among 100 random observers, but for the case with $1-b=0.6$ that compares better, while still providing lower statistics. A bias factor between $0.6$ and our baseline $0.8$ would result in a better match with these observed samples. 

The eROSITA numbers of detected clusters for the BigMD27-LC and BigMDPL-LC, with $1-b=0.8$ and $M_{500}\geq 5 \times 10^{13}$~$h_\mathrm{cosmo}^{-1}$~M$_\mathrm{sun}$, compares best with previous predictions from \cite{2012MNRAS.422...44P} and \cite{2014A&A...567A..65B}, respectively, which obtained $9.3\times10^4$ and $11.3\times10^{4}$ clusters, respectively, for the same observation time and sky coverage (0.658 of the sky, obtained excising $\pm20^\circ$ from the Galactic Plane; see \citealp{2010SPIE.7732E..0UP} and \citealp{2012MNRAS.422...44P}). In fact, the cosmological parameters adopted by \cite{2012MNRAS.422...44P} are close to our WMAP-7 cosmology, while those adopted by \cite{2014A&A...567A..65B} are closer to our \emph{Planck} cosmology. Note that both \cite{2012MNRAS.422...44P} and \cite{2014A&A...567A..65B} did not use a mass cut, as we do here, but rather a redshift-dependent count-rate cut that corresponds to a fiducial mass cut but additionally accounts for scatter effects and for the fact that, in real observations, one cannot perform a mass cut. There are other differences, of course, as the underlying adopted scaling relations, that we do not investigate further. 

\begin{figure*}
\centering
\includegraphics[width=.49\textwidth]{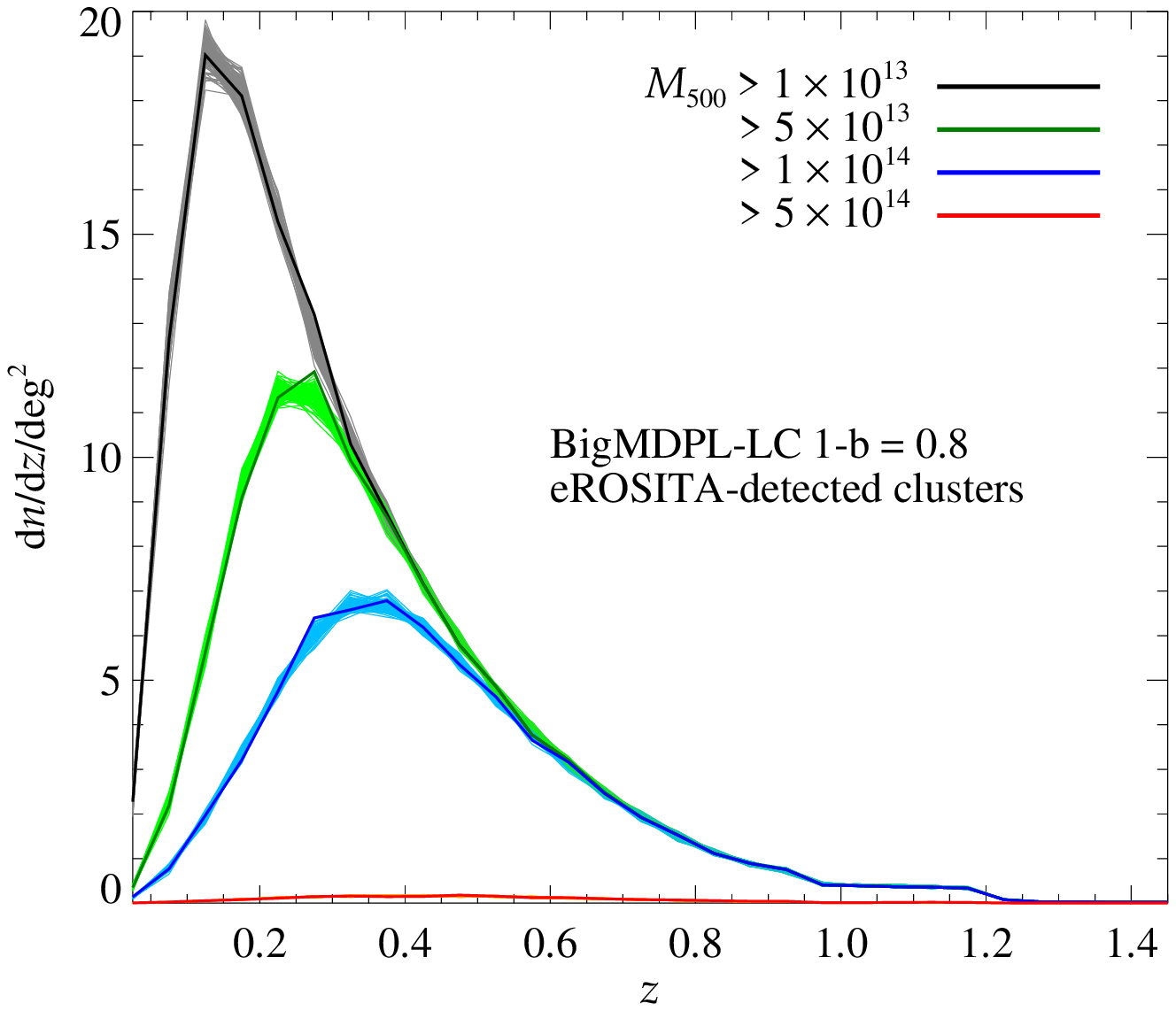}
\includegraphics[width=.49\textwidth]{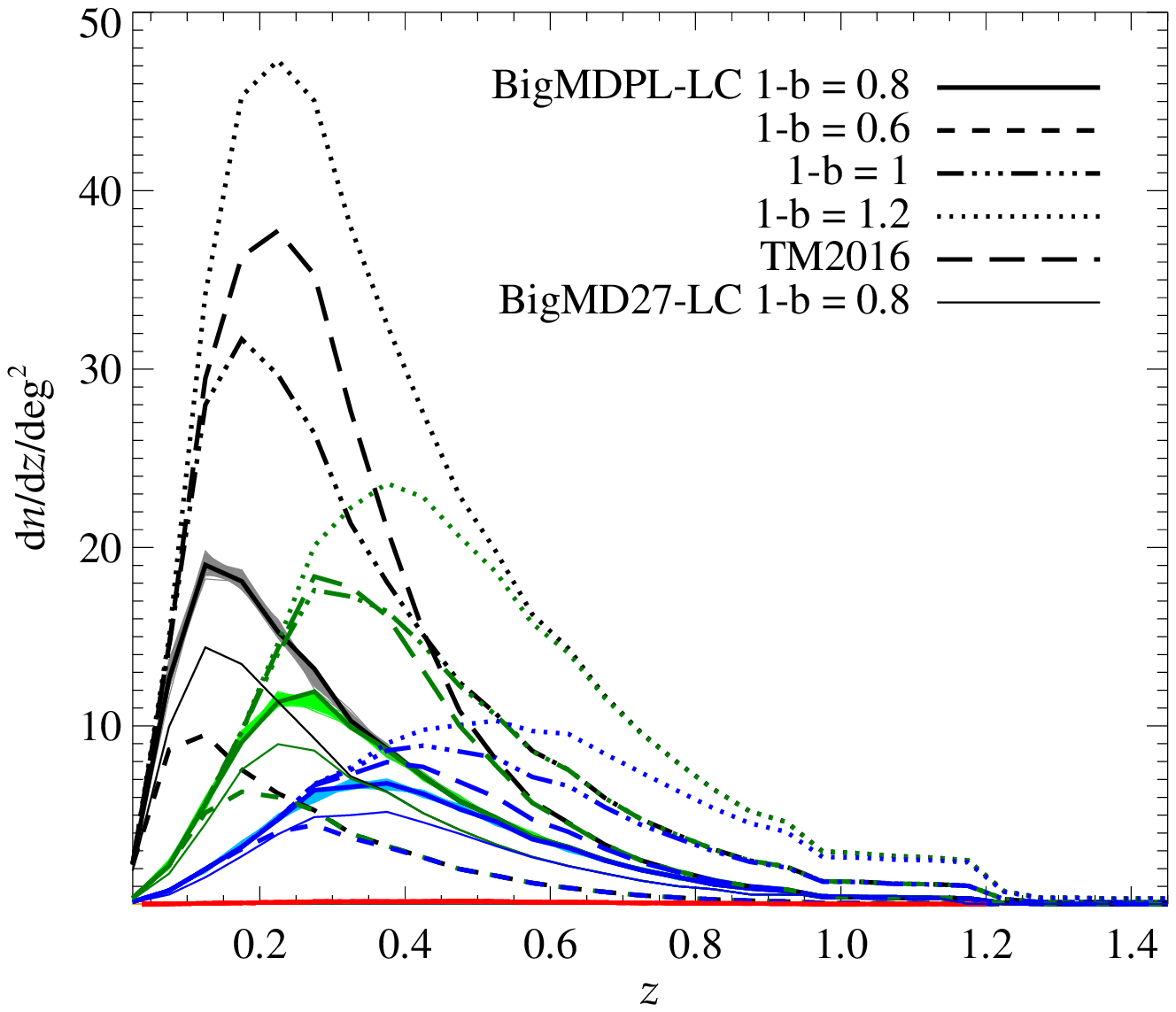}
\caption{\label{fig:num_halos_LC_2} Same as the right panel of  Figure~\ref{fig:num_halos_LC_1} but with linear y-scales. The left panel shows an enlarged view of the differences in the eROSITA-detected samples in the lower mass cuts in units of $h_\mathrm{cosmo}^{-1}$~M$_\mathrm{sun}$, while the right panel shows the differences among different phenomenological models and cosmologies using different mass cuts (colour-coded as in the left panel). Note in both panels the artefacts appearing at $z \gtrsim 0.94$ due to the limited number of available snapshots in the original simulation at high redshifts (see Section~\ref{sec:2.1}).}
\end{figure*}

Let us examine the statistics for the eROSITA-detected population of clusters in the different light-cones. It is particularly striking that despite the very different underlying assumptions in the different light-cones, the number of detected objects above the highest mass cut of $M_{500}\geq 5 \times 10^{14}$~$h_\mathrm{cosmo}^{-1}$~M$_\mathrm{sun}$ is very close. In the case of the BigMDPL-LC catalogues for the different phenomenological models, the variations are within the RMS of the 100 random observers of the $1-b=0.8$ case. The obvious effect of changing the value of $1-b$ is to assign, at fixed redshift, higher or lower fluxes at any given mass. This means that objects at higher redshift and/or with lower masses become detectable or undetectable changing $1-b$ to higher or lower values, respectively. However, the fact that we impose a lower mass cut when calculating these numbers together with the fact that the higher the mass the lower the number of galaxy clusters in our Universe, result in having smaller and smaller variations among different light-cones going to higher and higher mass cuts. It is also important to note that the higher the mass cut the less the effect of the 50-counts limit becomes: i.e., almost all clusters with $M_{500}\geq 5 \times 10^{14}$~$h_\mathrm{cosmo}^{-1}$~M$_\mathrm{sun}$ will be detected by eROSITA, almost independently of redshift and of flux variations due to the $1-b$ value, as we will see below. 

Another interesting note to make is that the BigMDPL-LC-TM2016 light-cone, with $1-b =0.8$ adopting the \protect\cite{2016MNRAS.463.3582M} $M_{500}-T$ relation, that is characterised by a shallower luminosity-mass relation with respect to our baseline, results in significantly boosting the number of detectable clusters at lower masses. It is difficult to say at this stage -- assuming our detection criterium was close to the true selection function of eROSITA, which is very likely not the case for low masses and high redshift -- what counts are reasonable to expect for very low mass cuts. However, recently \cite{2017MNRAS.469.3738S} found that galaxy groups with masses $M_{500} < 10^{14}$~$h_{0.7}^{-1}$~M$_\mathrm{sun}$ exhibit a significantly steeper luminosity-mass relation compared to clusters at higher masses. This is definitively something that can be implemented in our approach and whose effect should be investigated in future releases of our light-cones.

A final note on the results of Table~\ref{tab:halo_numbers} regarding the variation among the 100 random observers for BigMDPL-LC with $1-b=0.8$. Apart from the case of the HIFLUGCS flux cut where the RMS variations are of 11\%, the other cases show a remarkably low variation due to the fact that we are considering cumulative numbers integrating over all redshifts. We will see in the following that this is not the case when looking at the redshift distribution of the eROSITA-detected clusters. Note that our baseline light-cone, BigMDPL-LC with $1-b=0.8$, does not lie exactly on the average value of the 100 random observers, which was to be expected given that a particular observer was pre-selected according to the total number of objects present in the light-cones in the different redshift shells (see Section~\ref{sec:2.1}).

Figure~\ref{fig:num_halos_LC_1} shows the number of clusters against redshift, above different mass cuts, in the 100-random-observers light-cones (BigMDPL-LC with $1-b=0.8$). The left panel shows the total number of objects present in the light-cones together with the objects detectable by eROSITA, while the right panel shows a zoomed-in version of the objects detectable by eROSITA. In both panels the thicker lines represent the observer that we chose to be our reference one, which lies close to, but not exactly, on the average of the 100 random observers (again, to be expected given that a particular observer was pre-selected according to the total number of clusters present in the light-cones in the different redshift shells; see Section~\ref{sec:2.1}). Note also the appearance of artefacts at $z \gtrsim 0.94$ due to the limited number of available redshift snapshots in the original simulation at high redshifts (see Section~\ref{sec:2.1}). We show the results for four mass cuts as in Table~\ref{tab:halo_numbers}. For the eROSITA-detected objects, the 10\% and 90\% quantiles in the distribution of the differences between the 100 observers and the baseline one, restricting ourselves to the redshift range without artefacts $z \leq 0.95$, are -3.8\% and 3.4\% for $M_{500} > 10^{13}$, -4.1\% and 4.1\% for $M_{500} > 5 \times 10^{13}$, -5.1\% and 4.3\% for $M_{500} > 10^{14}$, and -30.2\% and 12.1\% for $M_{500} > 5 \times 10^{14}$~$h_\mathrm{cosmo}^{-1}$~M$_\mathrm{sun}$, respectively.

Figure~\ref{fig:num_halos_LC_2} shows, on the left, another version of the right panel of Figure~\ref{fig:num_halos_LC_1} that better visualises the differences for different mass cuts in the number of eROSITA-detected clusters and the variation among the 100 random observers. The right panel of Figure~\ref{fig:num_halos_LC_2} shows how the cluster counts change when changing the underlying phenomenological model and when changing from \emph{Planck} to the WMAP-7 cosmology of BigMD27-LC. Two things are clear for this set of figures. First that the variation among the 100 random observers is typically smaller than the variation among the models considered or cosmologies - that is not to say that it is negligible as cosmic variance is a irreducible source of systematic uncertainty in cosmological studies \citep{2003ApJ...584..702H,2011A&A...536A..95V}. Second that the variation among the different considered models and cosmologies are relatively small at low redshifts, below 0.2 to 0.4 depending on the mass cut, and are smaller and smaller when using higher and higher mass cuts. In fact, in samples with objects with $M_{500} > 5 \times 10^{14}$~$h_\mathrm{cosmo}^{-1}$~M$_\mathrm{sun}$, the variations among the different phenomenological models are basically indistinguishable from the baseline $1-b=0.8$ at all redshifts (for the reasons explained above), and the case of the WMAP-cosmology BigMD27-LC lies just below the envelope of the 100 BigMDPL-LC random observers. Therefore, samples including lower mass objects and redshift tomography will be crucial in constraining scaling relations and cosmological parameters with eROSITA (see also \citealp{2012MNRAS.422...44P} and \citealp{2014A&A...567A..65B}).

%%%%%%%%%%%%%%%%%%%%%%%%%%%%%%%%%%%%%%%%%%%%%%%%%%%%%%%%%%%%%%%%%%%
%%%%%%%%%%%%%%%%%%%%%%%%%%%%%%%%%%%%%%%%%%%%%%%%%%%%%%%%%%%%%%%%%%%
\section{(Angular) Power Spectrum of Clusters of Galaxies}
\label{sec:5}

In this section we use our light-cones with clusters with $M_{500}\geq1\times10^{13}$~$h_\mathrm{cosmo}^{-1}$~M$_\mathrm{sun}$ to investigate the power spectrum of the X-ray emission of clusters of galaxies. We project onto the sky the X-ray surface brightness of the clusters in our light-cones (see item (v) of Section~\ref{sec:2.2}) and then analyse these mock sky maps. We use two different, but related, quantities: the \emph{power spectrum} $P(k)$ and the \emph{angular power spectrum} $C_l$.

%%%%%%%%%%%%%%%%%%%%%%%%%%%%%%%%%%%%%%%%%%%%%%%%%%%%%%%%%%%%%%%%%%%
\subsection{Comparison to the XBOOTES field analysis by Kolodzig et al.}
\label{sec:5.1}

Before making predictions for the eROSITA clusters, we compare to current observational results of \cite{2017MNRAS.466.3035K,2018MNRAS.473.4653K} on the surface brightness fluctuations of the XBOOTES 9-deg$^2$ field of \emph{Chandra} \citep{2005ApJS..161....1M,2005ApJS..161....9K}. We follow closely \cite{2017MNRAS.466.3035K}, and compute the one-dimensional \emph{power spectrum} $P(k)  =  \langle |\widehat{\delta F}(k)|^2 \rangle$ as 
\begin{equation}
\widehat{\delta F}(\vec{k}) =  \dfrac{1}{\sqrt{\Omega}} \; \sum_{i,j}  \,\delta F(\vec{r}_{i,j}) \,\exp\left( -2\pi\,\mathrm{i}\; \vec{r}_{i,j} \cdot \vec{k}  \right) \, ,
\end{equation}
where $\delta F$ represents the surface brightness fluctuations of a given image (constructed by subtracting from each pixel the average value of all pixels in that image), $\vec{r}_{i,j}$ is the position of a pixel in the image, and $\Omega$ the solid angle of the image. In this way, having a $2\pi$ factor in the exponent, the angular scale $r$ is related to the frequency as $r=k^{-1}$. In calculating $P(k)$, the ensemble average $\langle |\widehat{\delta F}(k)|^2 \rangle$ is replaced with the average over all independent Fourier modes $\widehat{\delta F}(\vec{k})$ per angular frequency $k$ (see \citealp{2017MNRAS.466.3035K} for details). We use the FFTW library (version 3.3.6) to compute the Fourier transform (\citealp{FFTW05}; http://www.fftw.org).

\begin{figure}
\centering
\includegraphics[width=.45\textwidth]{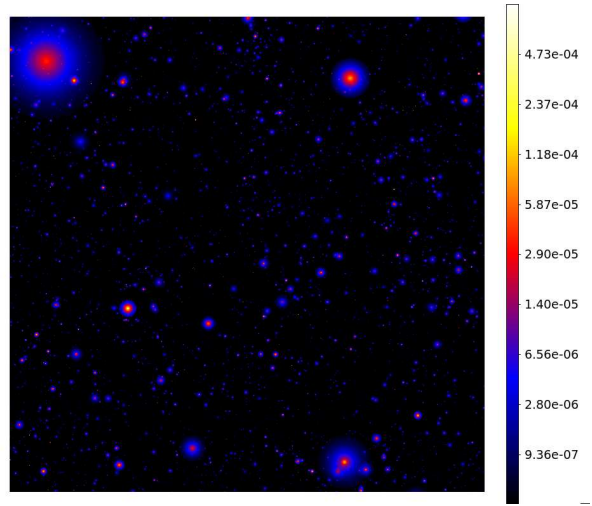}
\caption{\label{fig:maps_1} XBOOTES-like $3 \times 3$~deg$^2$ Cartesian sky map from our baseline light-cone BigMDPL-LC with $1-b=0.8$ for all clusters with $M_{500}\geq1\times10^{13}$~$h_{68}^{-1}$~M$_\mathrm{sun}$. This sky patch is our reference case as median representative of 30 random of such sky patches (see text for details). The logarithmic colour-coding represents the \emph{Chandra} count rate.}
\end{figure}

\begin{figure*}
\centering
\includegraphics[width=.91\textwidth]{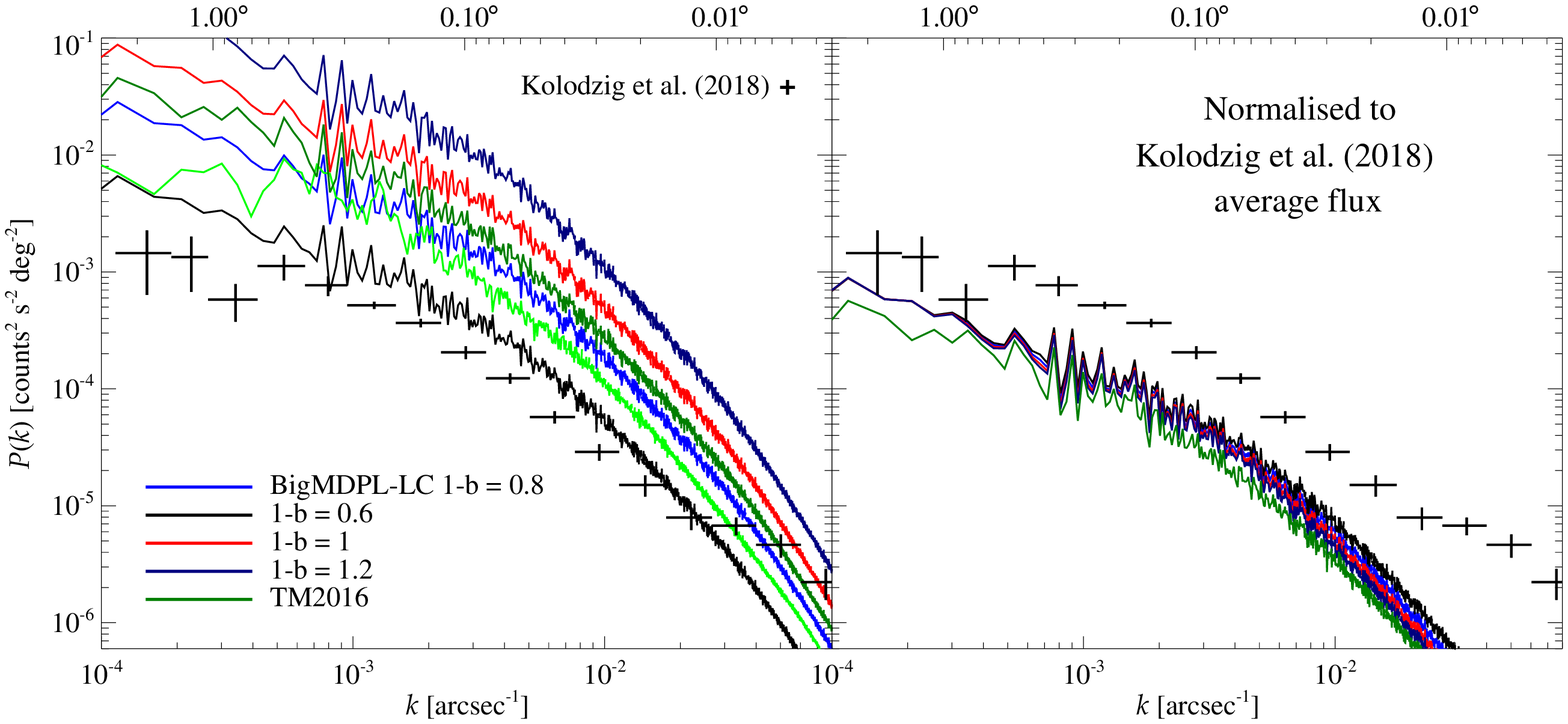}
\includegraphics[width=.91\textwidth]{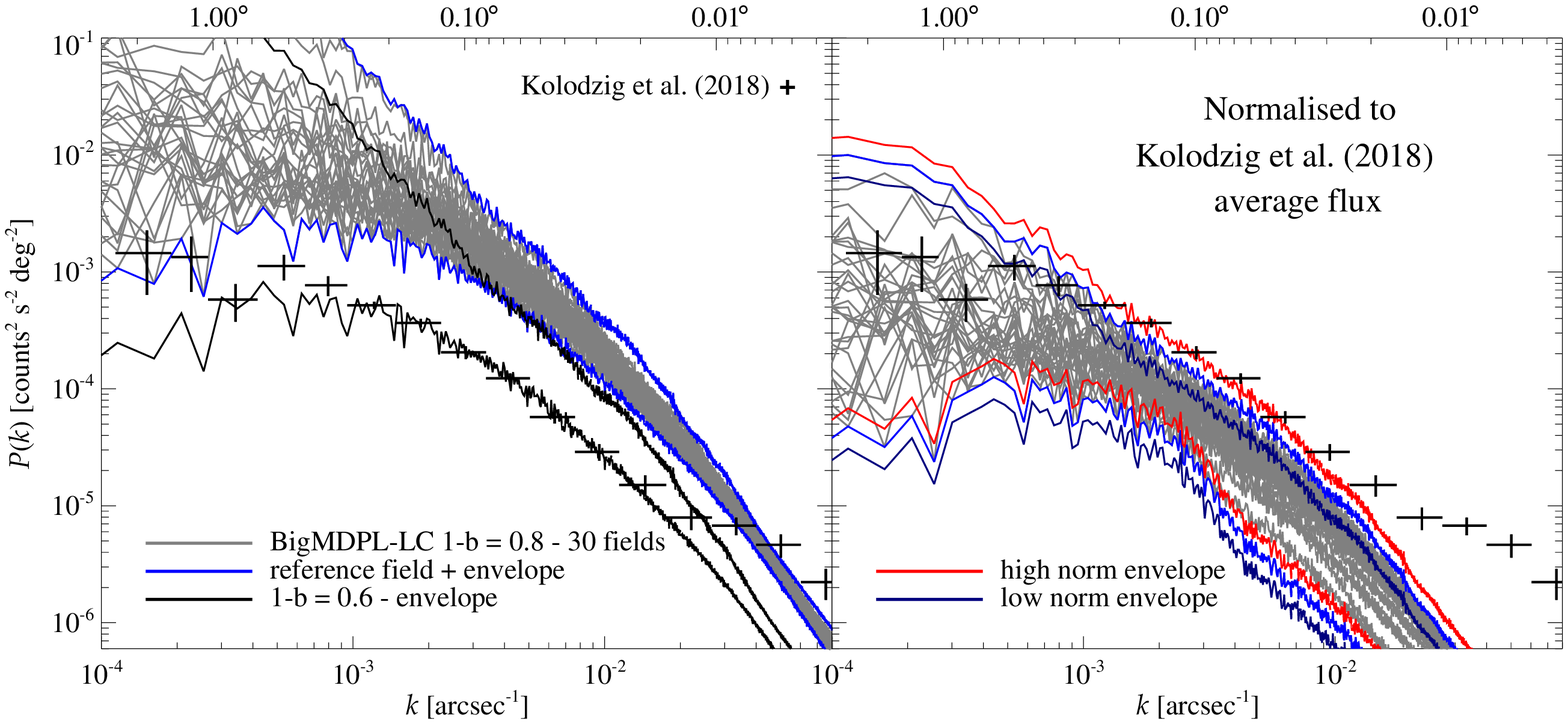}
\includegraphics[width=.91\textwidth]{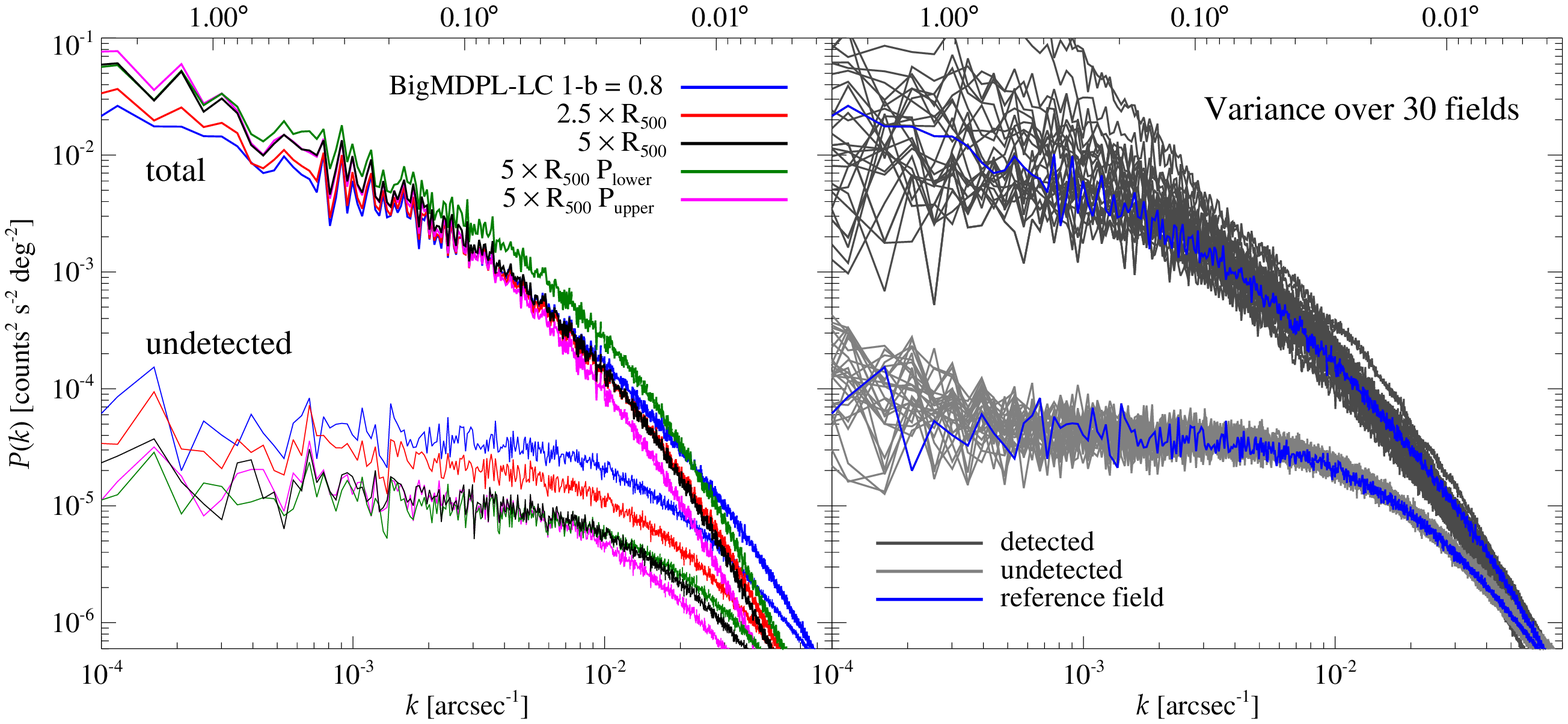}
\caption{\label{fig:pk} Comparison of the power spectra of the different phenomenological models of BigMDPL-LC with the data of \protect\cite{2018MNRAS.473.4653K} for the \emph{resolved and unresolved} cluster component (photon-shot-noise and AGN-shot-noise subtracted, and PSF-corrected) of the XBOOTES \emph{Chandra} field (top panels). Variance among 30 random mock XBOOTES sky patches (middle panels). Investigation of the effect of different integration radii and underlying pressure profiles, and of the variance in the detected and undetected cluster population (bottom panels) using an XBOOTES-like flux limit to separate the two populations \protect\citep{2017MNRAS.466.3035K}. See main text for details.}
\end{figure*}

For each of our light-cones, we calculate the \emph{Chandra} count rate in the $0.5-2$~keV energy range by using the response functions as obtained from the CIAO package (v4.7; CALDB v4.6.9; \citealp{2006SPIE.6270E..1VF}) and assuming a hydrogen column density of $10^{20}$~atom~cm$^{-2}$ for the photoelectric absorption as appropriate for the XBOOTES field (see \citealp{2017MNRAS.466.3035K} for details). We then use the projected flux surface brightness of each of our clusters (see item (v) of Section~\ref{sec:2.2}) and renormalise it to the \emph{Chandra} count rate to obtain projected count-rate sky maps. We construct many Cartesian (plate-carr\'ee) maps of random patches of the sky from our light-cones, with the same solid angle and resolution (pixel size of 0.492$''$) used by \cite{2018MNRAS.473.4653K}, that are our images in the sense of the previous paragraph. We show one of these sky maps in Figure~\ref{fig:maps_1}. For each of these patches, we eventually calculate the power spectrum $P(k)$ as explained above. Figure~\ref{fig:pk} contains six panels that compare the \emph{Chandra} count-rate power spectra from our light-cones with the data from \cite{2018MNRAS.473.4653K} for all (\emph{resolved and unresolved}) clusters in the XBOOTES field (photon-shot-noise and point-source-shot-noise subtracted, and PSF-corrected). 

Note that in this section, where we deal with small sky patches, we do not include in the discussion the effect of the WMAP-7 cosmology using BigMD27-LC as its effect would be indistinguishable from just another random sky patch taken from one of the BigMDPL-LC catalogues. In fact, because of the limited area of the considered sky patches, we would be anyway dominated by sample variance at most angular scales.

The top panels of Figure~\ref{fig:pk} show $P(k)$ of the BigMDPL-LC for different phenomenological models for our reference patch of the sky chosen to be the median between 30 random mock XBOOTES patches (see Figure~\ref{fig:maps_1} and below). The top-left panel shows the BigMDPL-LC power spectra against the \cite{2018MNRAS.473.4653K} data, while the top-right panel shows the same power spectra renormalised to the average (resolved and unresolved cluster) flux in the $0.5-2$~keV energy range estimated by \cite{2018MNRAS.473.4653K} of $5.1 \times 10^{-13}$~erg~cm$^{-2}$~s$^{-1}$~deg$^{-2}$. The middle panels of Figure~\ref{fig:pk} show the power spectra of 30 random sky patches from the baseline BigMDPL-LC with $1-b=0.8$, and the middle-right panel show these power spectra renormalised to the average flux estimated by \cite{2018MNRAS.473.4653K}. We believe that the 30 random patches of the sky are representative of the sample variance as the median, 10\% and 90\% quantiles of these power spectra for $k \gtrsim 3 \times 10^{-3}$~arcsec$^{-1}$ are stable already after about 10 realisations, while the variations at lower $k$, so at larger scales, are anyway dominated by large and bright nearby objects. The blue lines in the middle panels represent the chosen reference patch, being the median of the 30 random patches, and the upper and lower envelope. 

We are not concerned here in exactly reproducing the \cite{2018MNRAS.473.4653K} results, but rather to show that our MultiDark mock light-cones compare well with state-of-the-art measurements. We note that the $P(k)$ of the BigMDPL-LC with $1-b=0.6$ compares well with the \cite{2018MNRAS.473.4653K} measurement without any need of renormalisation, particularly when considering the lower envelope of the 30 random sky patches for this case, shown in black in the middle-left panel of Figure~\ref{fig:pk}. We are also pleased to note, looking at the middle-right panel of Figure~\ref{fig:pk}, that the ensemble of the renormalised power spectra of the 30 random fields compare decently with the measurement (but for $k \gtrsim 2 \times 10^{-2}$~arcsec$^{-1}$ that we will discuss below), while slightly undershooting it. The correct renormalisation factor is affected by significant uncertainty as the XBOOTES average (resolved and unresolved cluster) flux in the $0.5-2$~keV energy range of $5.1 \times 10^{-13}$~erg~cm$^{-2}$~s$^{-1}$~deg$^{-2}$ is obtained by summing the value for the resolved objects of $2.0~\pm~0.6 \times 10^{-13}$~erg~cm$^{-2}$~s$^{-1}$~deg$^{-2}$ -- which has a rather poor accuracy due to the low number of resolved cluster ($\sim$40) --  to that of the unresolved ones of $3.1 \times 10^{-13}$~erg~cm$^{-2}$~s$^{-1}$~deg$^{-2}$ \citep{2018MNRAS.473.4653K}. The latter also suffers from large uncertainties due to the fact that it is estimated with the X-ray luminosity function of \cite{2016A&A...592A...2P}, which is based on bright ($\gtrsim4\times10^{-14}$~erg~cm$^{-2}$~s$^{-1}$), mostly luminous ($\gtrsim10^{43}$~erg~s$^{-1}$) clusters and has to be extrapolated towards very small fluxes and luminosities. We try to convey this uncertainty in the middle-right panel of Figure~\ref{fig:pk}, where we show how the lower and upper envelope of the 30 random sky patches change when adopting a normalisation factor of $4.1$ and $6.1 \times 10^{-13}$~erg~cm$^{-2}$~s$^{-1}$~deg$^{-2}$ in dark blue and red, respectively. We chose these last two numbers just by assuming about $\pm20$\% variation in XBOOTES average flux quoted by \cite{2018MNRAS.473.4653K} which seems very plausible given the above discussion.

With such a small cluster sample and sky patch, we cannot identify the correct answer for the best match with the data among an overall normalisation related to the underlying cosmology, to the adopted ICM modelling, to the XBOOTES-field analysis, or a combination of some, or all, of these effects. At any rate, it is important to stress that the prediction for the cluster population power spectrum fall short of the measurement at small scales $k \gtrsim 2 \times 10^{-2}$~arcsec$^{-1}$. This was already noted by \cite{2017MNRAS.466.3035K} and could be due to several factors. On the measurement side, this could be due to an underestimation of the point-source shot-noise contribution at these small scales. However, this seems less likely given the good agreement of the estimated point-source shot-noise with predications from the literature \citep{2018MNRAS.473.4653K}. On the modelling side, this could be due to an increased flux of small mass halos that, however, seems unlikely to be enough given the little difference among our phenomenological models, particularly when looking at BigMDPL-LC-TM2016 that we already discussed in Section~\ref{sec:4} to be quite extreme in this sense. Alternatively, as this is the angular regime where we start to resolve the inner structure of clusters, this could be due to inhomogeneities in the ICM that our model does not include, in particular, for example, to ICM clumpiness in clusters outskirts (A.~Kolodzig, private communication). Obviously, a combination of these factors could also explain the discrepancy.

In the bottom-left panel of Figure~\ref{fig:pk}, we investigate the role of the integration radius for our baseline light-cone and reference sky patch. Our mock light-cones generally include quantities only up to $R_{500}$, which can be insufficient to get a complete idea of the power spectrum. We show what happens when increasing the integration radius to $2.5 \times R_{500}$ and $5 \times R_{500}$, noting that, while both cases are quite different than the $R_{500}$ power spectrum, they do not differ much among themselves suggesting that $5 \times R_{500}$ is a good integration radius for such studies. The regime beyond this limit is much more speculative given that the \emph{Planck} measurements of the pressure profiles, on which our light-cones are based, do not go that far out. Note that the increase of integration radius corresponds to a steeper fall of the power spectra at small (large) angular scales ($k$) as we start resolving the inner structure of the halos. In fact, the halo power spectrum is typically characterised by a cut-off at small scales corresponding to some characteristic scale of the profile under consideration (see, e.g., \citealp{2006PhRvD..73b3521A}).

\begin{figure}
\centering
\includegraphics[width=.49\textwidth]{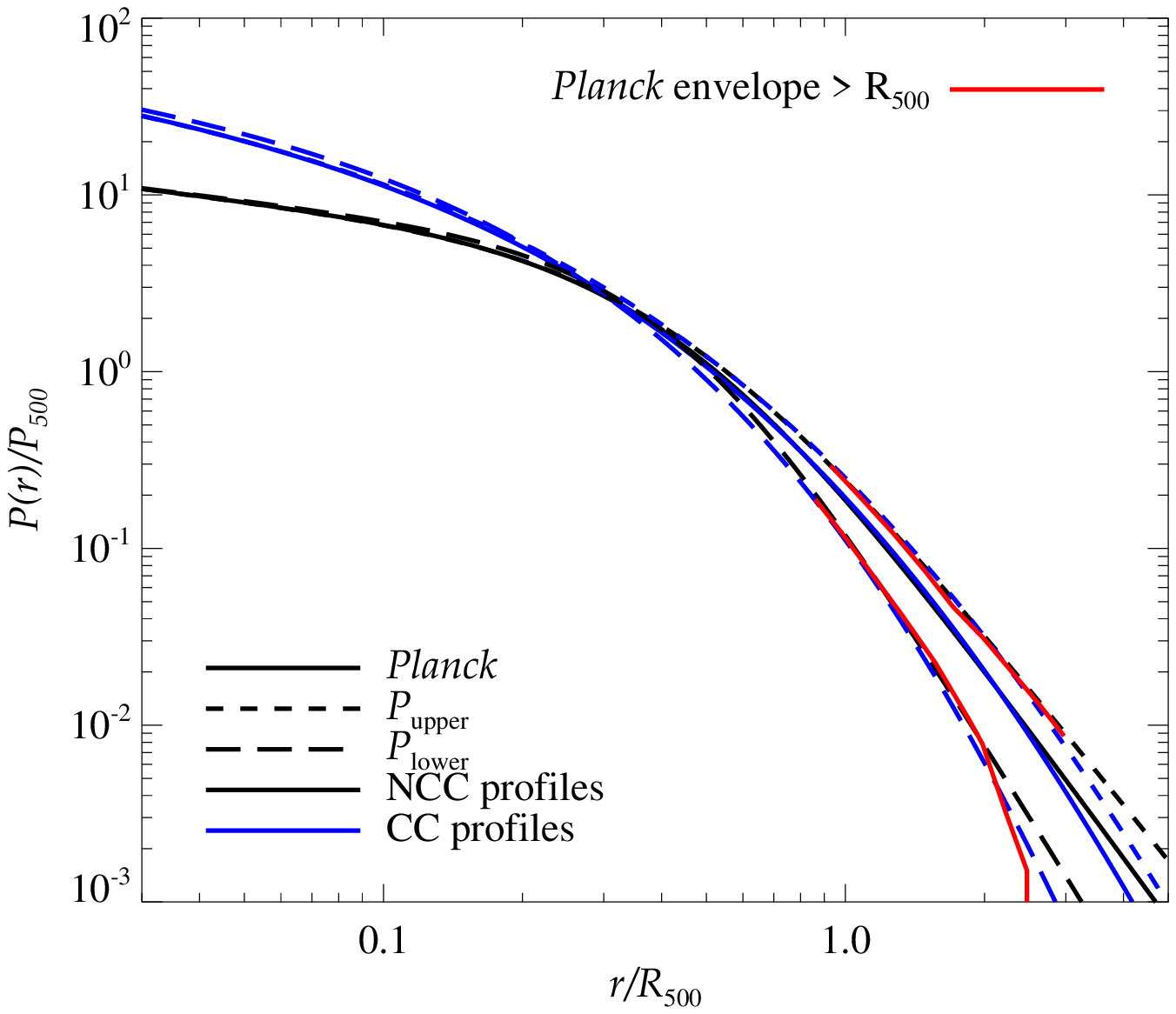}
\caption{\label{fig:P_envelope} Pressure profiles from which our phenomenological model is built. We show our baseline choice of the \emph{Planck} profiles for non-cool-core (NCC; $T_\mathrm{drop}=0.8$, $1$) and cool-core (CC; $T_\mathrm{drop}=0.4$, $0.6$) clusters \protect\citep{2013A&A...550A.131P}, together with the \emph{ad hoc} pressure profiles that we constructed modifying the GNFW parametrisation of \protect\cite{2013A&A...550A.131P} to sample the upper and lower envelope of the \emph{Planck} pressure profiles at $\gtrsim R_{500}$ (red lines). As discussed in the text, for the case of $P_\mathrm{lower}$ (long-dashed lines), this incidentally resulted in modifying the pressure also at small radii, generating a mild boost with respect to the baseline \emph{Planck} profiles.}
\end{figure}

In order to investigate the effect of different profiles, in the bottom-left panel of Figure~\ref{fig:pk}, we also include the power spectra for two more cases up to $5 \times R_{500}$. We built from scratch two more BigMDPL-LC catalogues (with $1-b=0.8$) that instead of using the mean \emph{Planck} cool-core and non cool-core pressure profiles, use these two profiles as located at the upper (indicated as $P_\mathrm{upper}$) and lower (indicated as $P_\mathrm{lower}$) envelope of the \emph{Planck} pressure profile measurements (figure~4 of \citealp{2013A&A...550A.131P}). The effect on the total signal of all clusters clearly follows what was discussed above regarding the small-scale cut-off of the halo power spectrum (e.g., \citealp{2006PhRvD..73b3521A}). $P_\mathrm{upper}$, implying a larger emission with respect to our baseline, results in a reduced power at small scales, while $P_\mathrm{lower}$, implying a smaller emission with respect to our baseline, results in an increased power. Note also that the case with $P_\mathrm{lower}$ gives an increase in power also at large scales. This is due to the fact that we constructed artificially the cool-core and non cool-core pressure profiles for our $P_\mathrm{upper}$ and $P_\mathrm{lower}$ cases by manually changing the parameters in the \emph{Planck} generalised Navarro-Frenk-White (GNFW) parametrisation of the pressure profiles. For the case of $P_\mathrm{lower}$, this incidentally resulted in modifying the pressure also at small radii, as can be seen from Figure~\ref{fig:P_envelope}. We decided to keep this case without modifying the inner part of the profile as it is illustrative to look at the effect onto the power spectrum discussed in this section, and onto the angular power spectrum of the eROSITA clusters that will be discussed in the next section.

In the bottom-left panel of Figure~\ref{fig:pk}, we also show the signal that would correspond to the undetected population of clusters, using as upper flux cut, in the $0.5-2$~keV range, $3 \times 10^{-14}$~erg~cm$^{-2}$~s$^{-1}$ \citep{2017MNRAS.466.3035K}. In these power spectra, two different effects can be appreciated. First, an overall different normalisation as with the increasing integration radius more and more objects are detected due to an overall increase of their total flux. This is also due to the fact that we use, for simplicity, the detection limiting flux cut without any correction for the integration radius. Second, a reduced (increased) power at small scales when adopting $P_\mathrm{upper}$ ($P_\mathrm{lower}$), but less pronounced than for the total power spectra as the previously mentioned effect is competing with the small-scale cut-off of the halo power spectrum.

In the bottom-right panel of Figure~\ref{fig:pk}, we show the variance among the 30 random sky patches mimicking the XBOOTES \emph{Chandra} field for our baseline BigMDPL-LC with $1-b=0.8$ dividing the power spectra in detected and undetected populations using the above mentioned flux cut of $3 \times 10^{-14}$~erg~cm$^{-2}$~s$^{-1}$ in the $0.5-2$~keV range \citep{2017MNRAS.466.3035K}. It is interesting to note the much smaller effect of sample variance on the power spectrum of the undetected population of clusters with respect to the detected one, as expected from a larger number for the former. Looking now back at all the panels of Figure~\ref{fig:pk}, the dramatic effect that sample variance can have when trying to extract global physical information on clusters using a small patch of the sky as that of XBOOTES is clear. While the largest scales ($\gtrsim$~10$'$; $k \lesssim 3 \times 10^{-3}$~arcsec$^{-1}$) seem hopeless as they are dominated by the variance among the largest, closest, brightest clusters in different sky patches, the smaller scales ($\lesssim$~2$'$; $k \gtrsim 10^{-2}$~arcsec$^{-1}$) and, in particular, the undetected population hold more discriminating power between different models if the measurement statistical and systematic (including the possible contribution of other sources as AGN and normal galaxies) errors can be kept at bay. While we argue that it is difficult to extract global physical information on clusters from small XBOOTES-like sky patches because of the sample variance, we are not doubting the usefulness of power spectrum analyses of a given sky patch in extracting information on the cluster population living in that particular patch. 

The conclusions of this section are that our mock light-cones compare well - with the mentioned limitations and uncertainties - with state-of-the-art measurements of the X-ray surface brightness fluctuations of the XBOOTES field \citep{2017MNRAS.466.3035K,2018MNRAS.473.4653K}, and that, in order to fully exploit the potential of such an approach to study the ICM of clusters and the composition of the cosmic X-ray background, it is paramount to overcome sample variance with all-sky surveys as the eROSITA one that we discuss in the next section.

%%%%%%%%%%%%%%%%%%%%%%%%%%%%%%%%%%%%%%%%%%%%%%%%%%%%%%%%%%%%%%%%%%%
\subsection{Predictions for the eROSITA all-sky survey}
\label{sec:5.2}

\begin{figure}
\centering
\includegraphics[width=.5\textwidth]{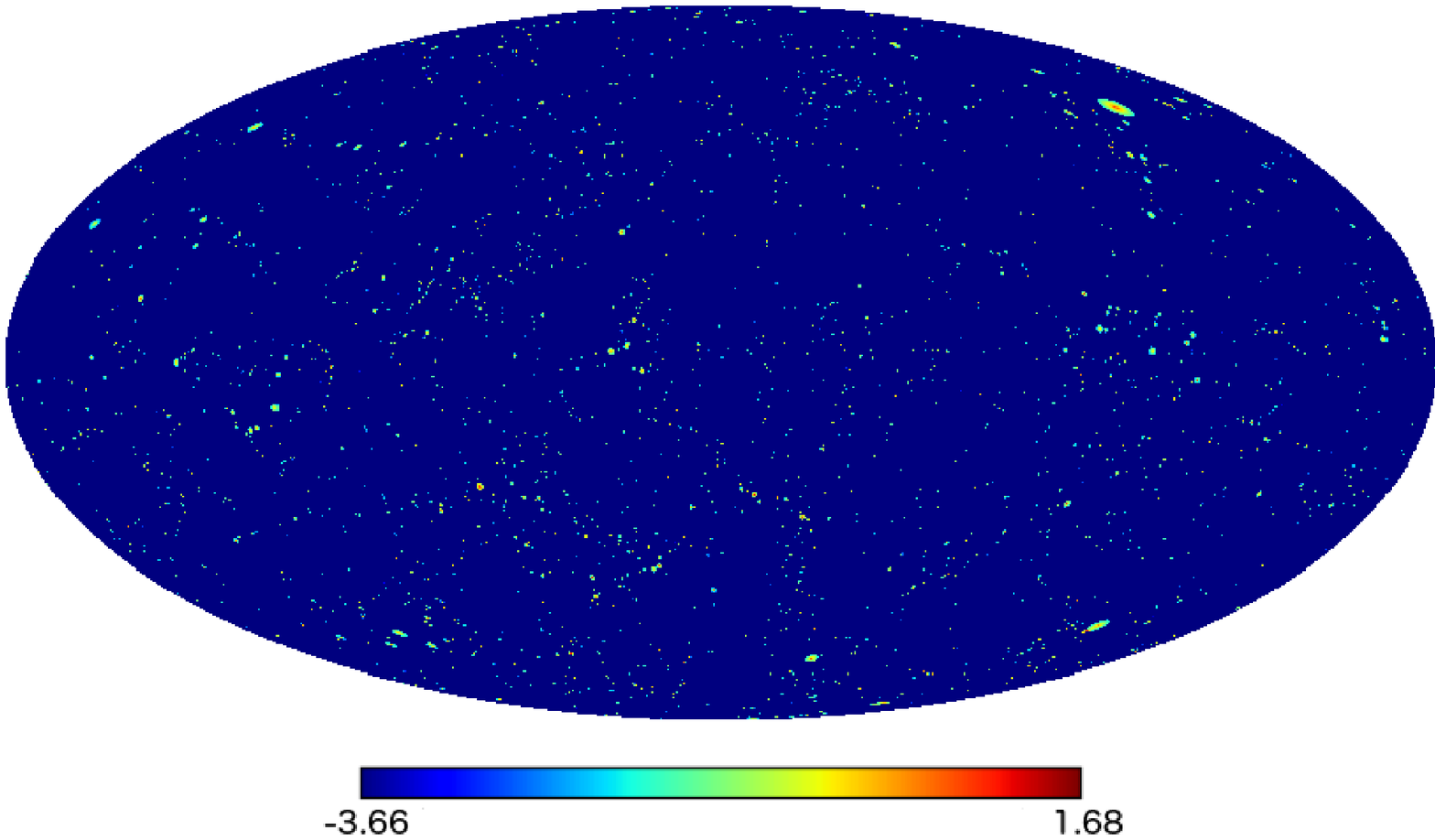}
\includegraphics[width=.5\textwidth]{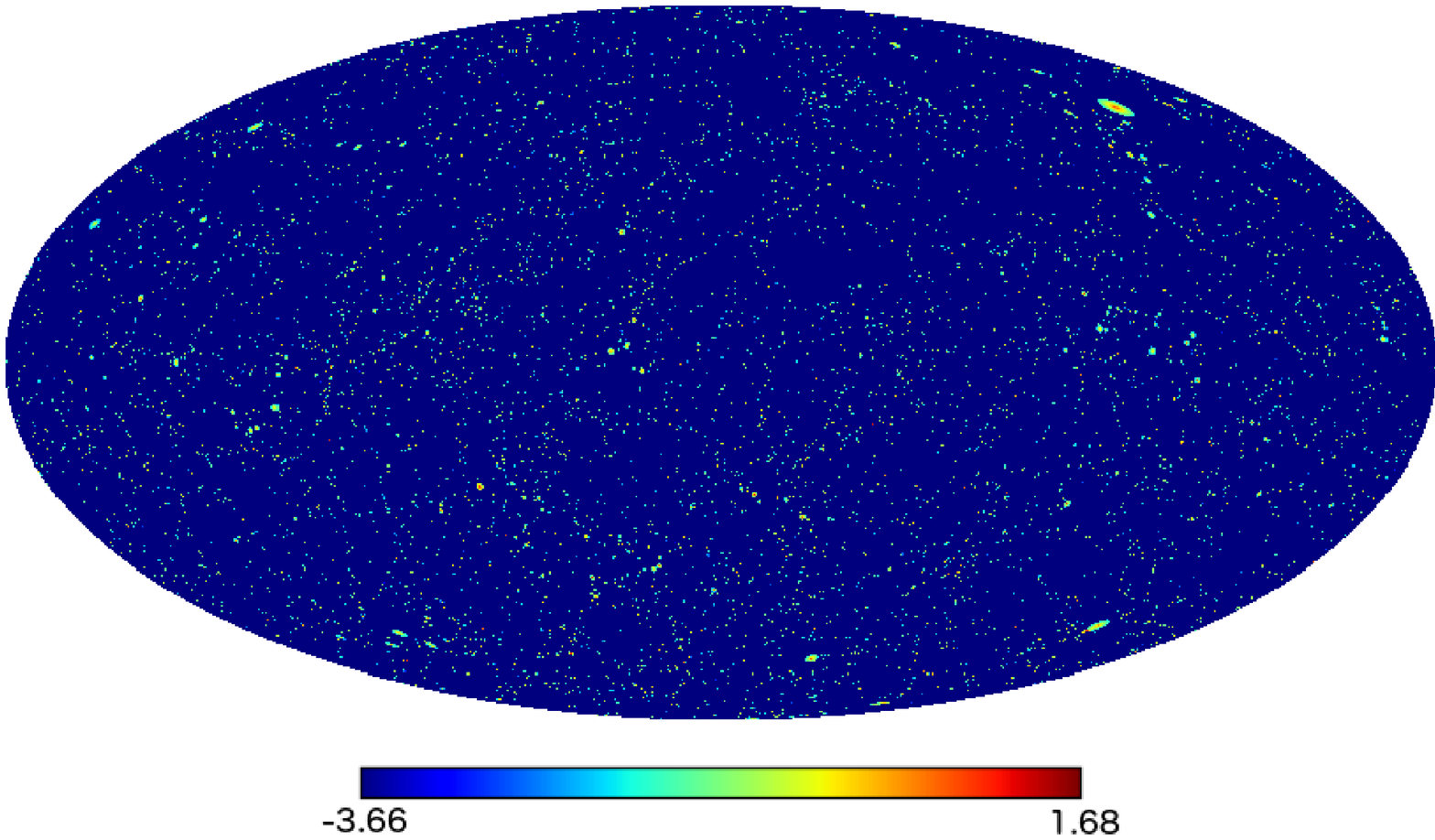}
\includegraphics[width=.5\textwidth]{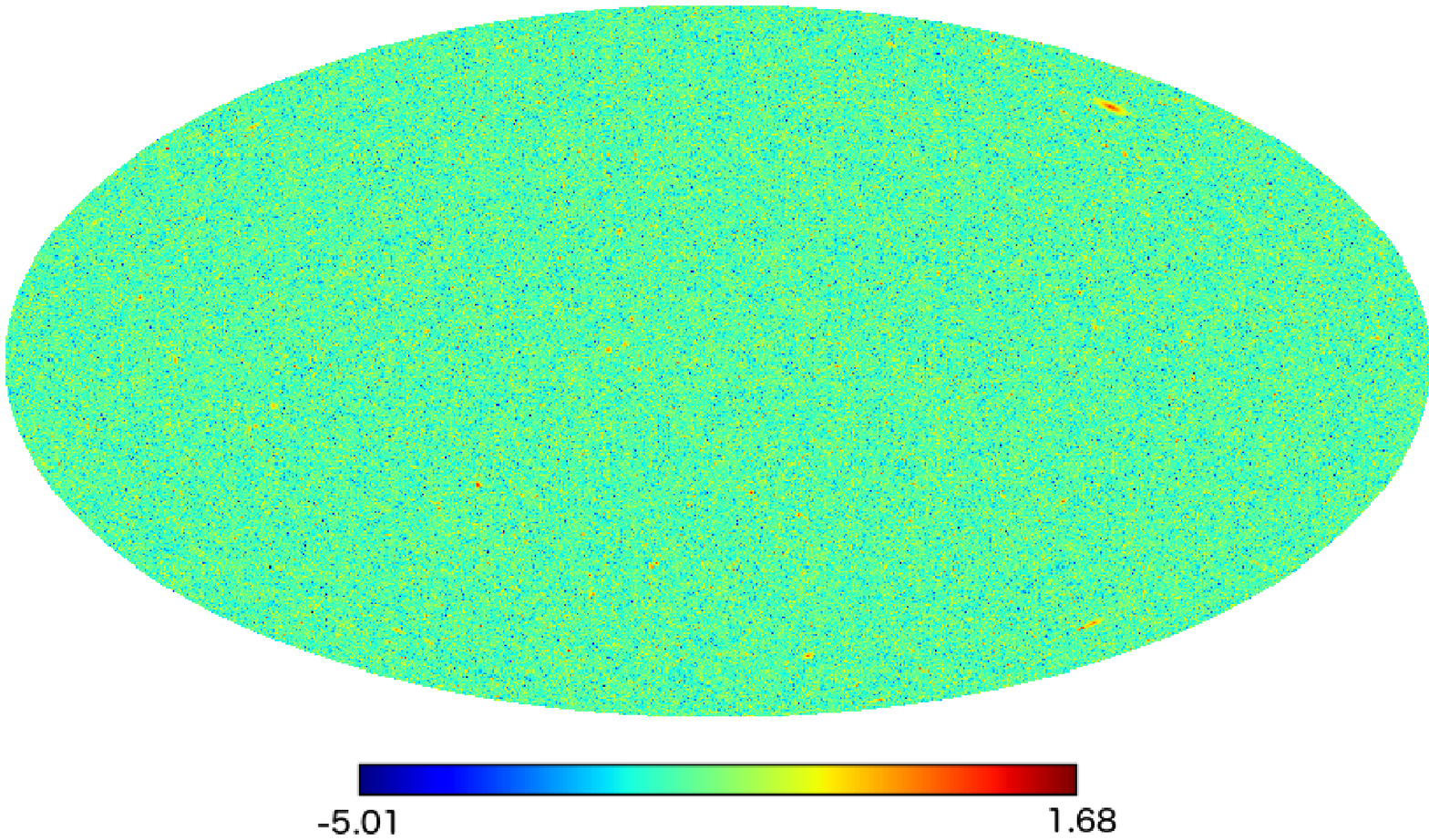}
\caption{\label{fig:maps_2} Simulated eROSITA count-rate all-sky HEALPix map ($N_\mathrm{side} = 512$) of our baseline light-cone BigMDPL-LC with $1-b=0.8$ for clusters with $M_{500}\geq1\times10^{13}$~$h_\mathrm{cosmo}^{-1}$~M$_\mathrm{sun}$. The numbers on the colour bar are log$_{10}$ of the counts rate. {\bf From top to bottom:} clusters at 
$z \leq 0.05$, clusters at $z \leq 0.1$, all clusters. For visualisation purposes, the void pixels in the images are set to the value of the pixel with the lowest count-rate.}
\end{figure}

\begin{figure*}
\centering
\includegraphics[width=.7\textwidth]{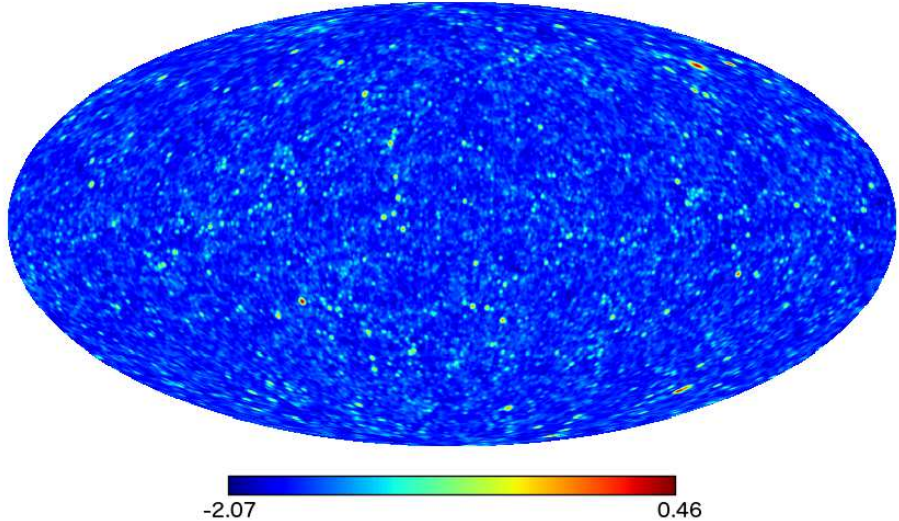}
\caption{\label{fig:maps_3} Smoothed simulated eROSITA count-rate all-sky HEALPix map ($N_\mathrm{side} = 512$) of our baseline light-cone BigMDPL-LC with $1-b=0.8$ for all clusters with $M_{500}\geq1\times10^{13}$~$h_\mathrm{cosmo}^{-1}$~M$_\mathrm{sun}$. The numbers on the colour bar are log$_{10}$ of the counts rate. This map is the same as the bottom one of Figure~\protect{\ref{fig:maps_2}} but smoothed with a Gaussian beam with a full-width half-maximum of $1^{\circ}$ to better visualise the large-scale structure. For visualisation purposes, the void pixels in the image are set to the value of the pixel with the lowest count-rate. We provide on-line at Skies \& Universes the fits file version of the high resolution image ($N_\mathrm{side} = 2048$).}
\end{figure*}

To make predictions for the eROSITA all-sky maps, we use the \emph{angular power spectrum} defined as (e.g., \citealp{2014A&A...568A..57H}):
\begin{equation}
C_l = \frac{1}{2 l +1} \sum_{l,m} | a_{l m} | ^2
\end{equation}
where $a_{lm}$ are the coefficients of the decomposition in spherical harmonics of a given image, in this case of a given all-sky map. We then use the HEALPix package (\citealp{2005ApJ...622..759G}; https://healpix.jpl.nasa.gov/) to calculate the angular power spectrum from our sky maps. In Figure~\ref{fig:maps_2} and \ref{fig:maps_3}, we show some example of eROSITA all-sky HEALPix maps for our baseline light-cone.

We generate, from our light-cones, all-sky maps by projecting all halos in our catalogues onto the sky. We choose to work with count-rate sky maps and, therefore, renormalise the projected flux surface brightness of each of our clusters (see item (v) of Section~\ref{sec:2.2}) to the eROSITA count rate. We generate all-sky maps with the HEALPix pixelisation scheme, i.e., equal-area pixels, where our standard resolution choice is $N_\mathrm{side} = 2048$ corresponding to a resolution of about $100$~arcsec to compare with the expected average PSF of the eROSITA all-sky survey of $30$~arcsec \citep{2012arXiv1209.3114M}. Our choice is to have a resolution as good as possible, so as close as possible to the expected eROSITA average, while maintaining a reasonably low computation time as we will be making this calculation many times.\footnote{In order to produce an all-skymap with the eROSITA average resolution of about $30$~arcsec, we would need to use $N_\mathrm{side} = 8192$, increasing the number of pixels from about $5\times10^{7}$, with $N_\mathrm{side} = 2048$, to about $8\times10^{8}$ with a significant increase in computation time. This is not prohibitive \emph{per se} as it would imply a computation time of about 1-2 days for an all-sky map on a normal desktop computer, but it is not important for the scope of this paper.} The corresponding angular power spectra, calculated with HEALPix as explained above and customarily plotted as $C_l (l+1) l / 2 \pi$, will be shown above $l = 10$, as larger scales are cut due to the exclusion $\pm20^\circ$ of the sky around the Galactic Plane \citep{2012MNRAS.422...44P}, up to the maximum resolved multipole corresponding to our chosen resolution.

\begin{figure*}
\centering
\includegraphics[width=.95\textwidth]{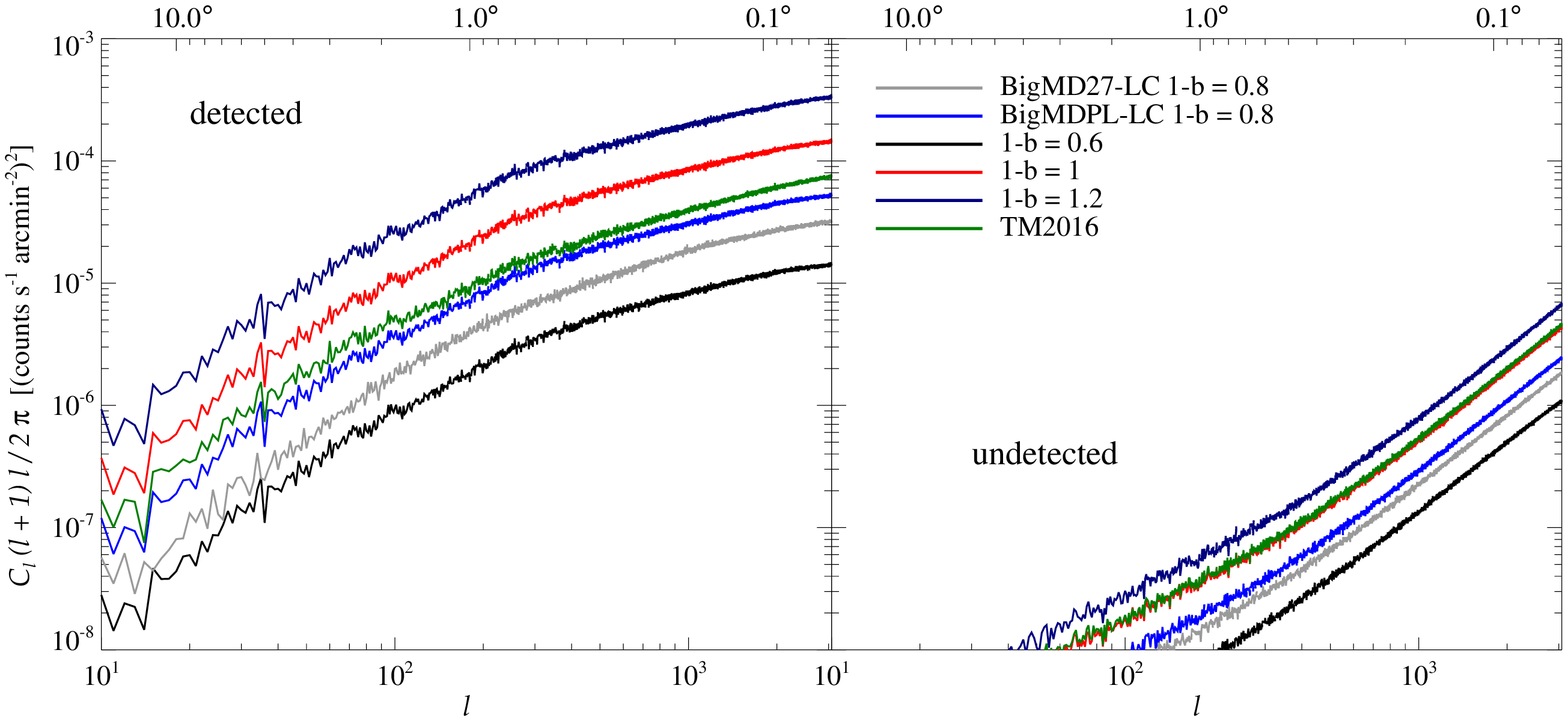}
\includegraphics[width=.95\textwidth]{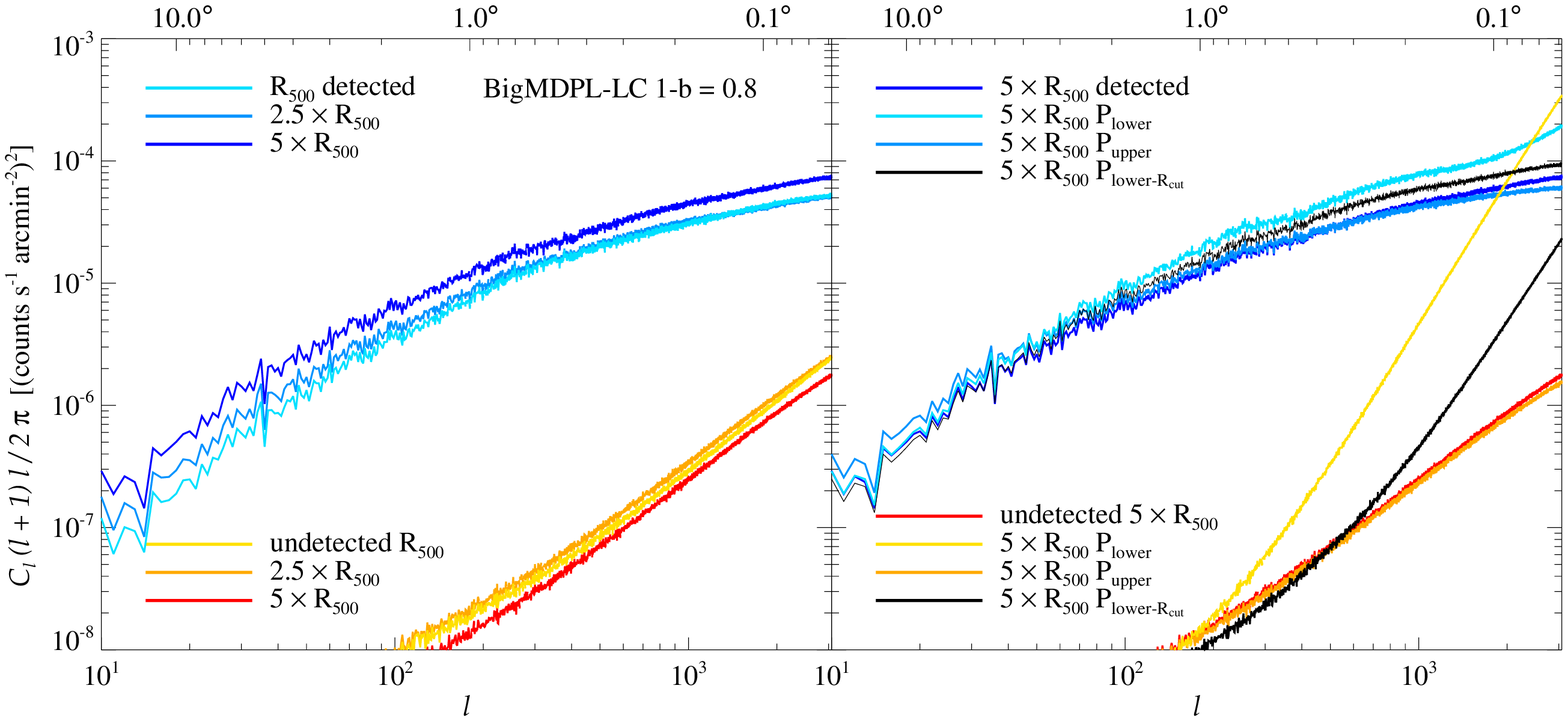}
\caption{\label{fig:cl} Prediction for the angular power spectra of the eROSITA all-sky survey cluster population. The top panels explore the differences in light-cones with different phenomenological models and underlying cosmologies, where the left panel shows the eROSITA-detected clusters and the right panel the undetected ones. The bottom panels explore, for our baseline BigMDPL-LC with $1-b=0.8$, the effect of increasing the integration radius (left panel) and of changing the underlying pressure profiles to sample the \emph{Planck} profiles envelope (right panel). The model dubbed $P_\mathrm{lower-R_{cut}}$ in the bottom-right panel refers to a pressure profile build by using the original \emph{Planck} pressure profiles for $R < 0.5 \times R_{500}$ and using $P_\mathrm{lower}$ otherwise. See main text for details. The top axes indicate the angular scale as $180^{\circ}/l$.}
\end{figure*}

\begin{figure}
\centering
\includegraphics[width=.49\textwidth]{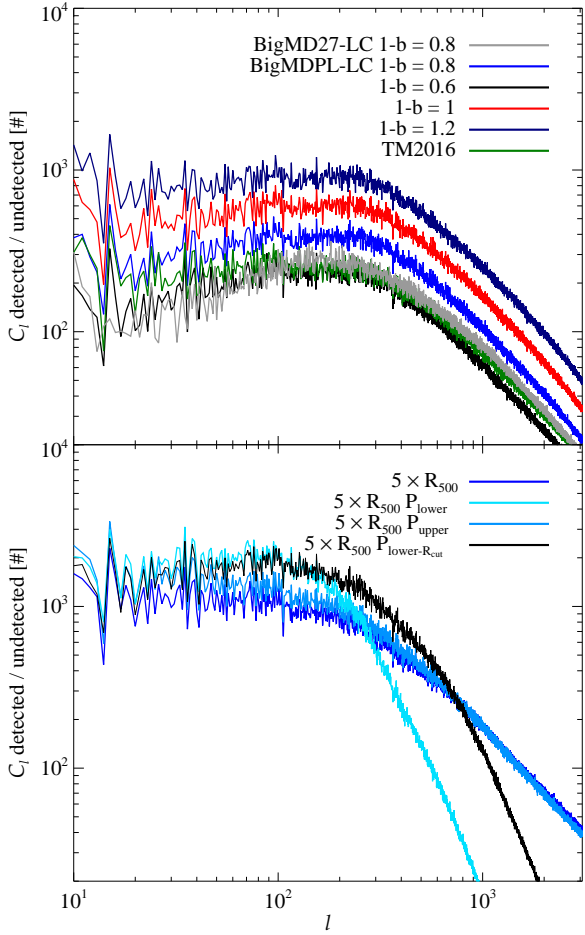}
\caption{\label{fig:cl_ratio} Ratio of the detected and undetected components of the angular power spectra of the eROSITA all-sky survey cluster population as from Figure~\ref{fig:cl}. The top panel shows the different phenomenological models and underlying cosmologies, while the bottom panel shows the effect of changing the underlying pressure profiles to sample the \emph{Planck} profiles envelope.}
\end{figure}

The top panels of Figure~\ref{fig:cl} show the all-sky angular power spectra from the different models of BigMDPL-LC and from the WMAP-7-cosmology  BigMD27-LC. The left panel shows the power spectra of the eROSITA-detected clusters, using as detection criteria the already mentioned threshold of $\geq$ 50 counts in 1.6~ks of observations (see Section~\ref{sec:4}), while the right panel shows the power spectra of the undetected clusters that would be part of the cosmic X-ray background. The detected objects dominate over the undetected ones, similarly to the $P(k)$ of the previous section but more clearly at all the angular scales considered here. However, the undetected angular power spectra sharply rise at large multipoles (small angular scales), so we could expect this component to be similar in magnitude to the detected component around $l=10^4$. This can be appreciated in the detected/undetected ratio shown in Figure~\ref{fig:cl_ratio}. Similarly to the $P(k)$ of the previous section, the most obvious difference between different phenomenological models is the overall normalisation of the signal. However, in this case, the difference in shape of the BigMDPL-LC with the harder mass-temperature relation of \cite{2016MNRAS.463.3582M} both at large and small scales, and on the undetected population, is clearly appreciable. In general, it appears that, for the analysed multipoles, the differences between different models and cosmologies are more pronounced for the detected population. This is not surprising given the large number of clusters that eROSITA should be able to detect with respect to current observational samples \citep{2012MNRAS.422...44P,2014A&A...567A..65B,2018arXiv180608652C}.

In the bottom panels of Figure~\ref{fig:cl}, we investigate again the effect of different integration radii and different underlying pressure profiles. The left panel shows the effect of increasing the integration radius from $R_{500}$ to $2.5 \times R_{500}$ and $5 \times R_{500}$. Differently from the $P(k)$ case of the previous section, here the cases of $R_{500}$ and $2.5 \times R_{500}$ lie roughly on top of each other both for the eROSITA-detected and undetected population for most multipoles. We could expect more differences at larger multipoles (smaller angular scales) where we could resolve more into the inner structure of most halos, similarly to what was shown in the previous section. The case with $5 \times R_{500}$ just results in a higher (lower) overall normalisation for the eROSITA-detected (undetected) clusters due to the fact that more clusters cross over the detection threshold of 50~counts (which, again, for simplicity, we do not correct for the integration radius).

The most interesting prediction is shown in the bottom-right panel of Figure~\ref{fig:cl} where we investigate the effect of changing the underlying pressure profiles from our baseline to $P_\mathrm{upper}$ and $P_\mathrm{lower}$, constructed as described in the previous Section~\ref{sec:5.1}, all integrating up to $5 \times R_{500}$. We remind the reader that $P_\mathrm{upper}$ and $P_\mathrm{lower}$ were obtained by manually changing the parameters in the \emph{Planck} GNFW parametrisation of the pressure profiles, and that the $P_\mathrm{lower}$ profiles incidentally resulted in modifying the pressure also at small radii as shown in Figure~\ref{fig:P_envelope}. Our baseline case and the $P_\mathrm{upper}$ case lie roughly on top of each other for most angular scales, but for very high multipoles (small angular scales) where we start to see the effect of resolving into the inner structure of halos discussed in the previous section in the context of the $P(k)$. On the contrary, the case of $P_\mathrm{lower}$ is quite interesting and illustrative of the power of this approach. The effect, barely noticeable in the bottom-left panel of Figure~\ref{fig:pk}, is dramatic here. The $P_\mathrm{lower}$ eROSITA-detected population is affected by a general mild boost in power, with respect to our baseline model, at most multipoles, with a somewhat more pronounced enhancement at larger multipoles (smaller scales) due to both the mildly higher emission at small radii in the $P_\mathrm{lower}$ pressure profiles, at parity of halo, with respect to our baseline model, and to the effect of resolving into the inner structure of halos. More importantly, the $P_\mathrm{lower}$ undetected population shows an extremely sharp rise toward larger multipoles (small scales) with respect to our baseline model due to an increasing number of halos dropping off from the eROSITA-detected population into the undetected one because of the steeper pressure profiles in the outskirts. This can also be seen in the detected/undetected ratio shown in Figure~\ref{fig:cl_ratio}. To further strengthen this point, we also show the build-up of this effect with a light-cone constructed using the original \emph{Planck} pressure profiles for $R < 0.5 \times R_{500}$ and $P_\mathrm{lower}$ otherwise that we call $P_\mathrm{lower-R_{cut}}$ in the legend of the bottom-right panel. As expected, the results for such a light-cone lie, both for the detected and undetected population, midway between the results of $P_\mathrm{lower}$ and those of the other two light-cones. This comparison beautifully shows all the discriminating power of an (angular) power spectrum analysis. However, the reader should keep in mind that different surface brightness profiles will likely result in different detection likelihoods that could reduce the strong differences in the $C_l$ trends discussed here.

\begin{figure}
\centering
\includegraphics[width=.49\textwidth]{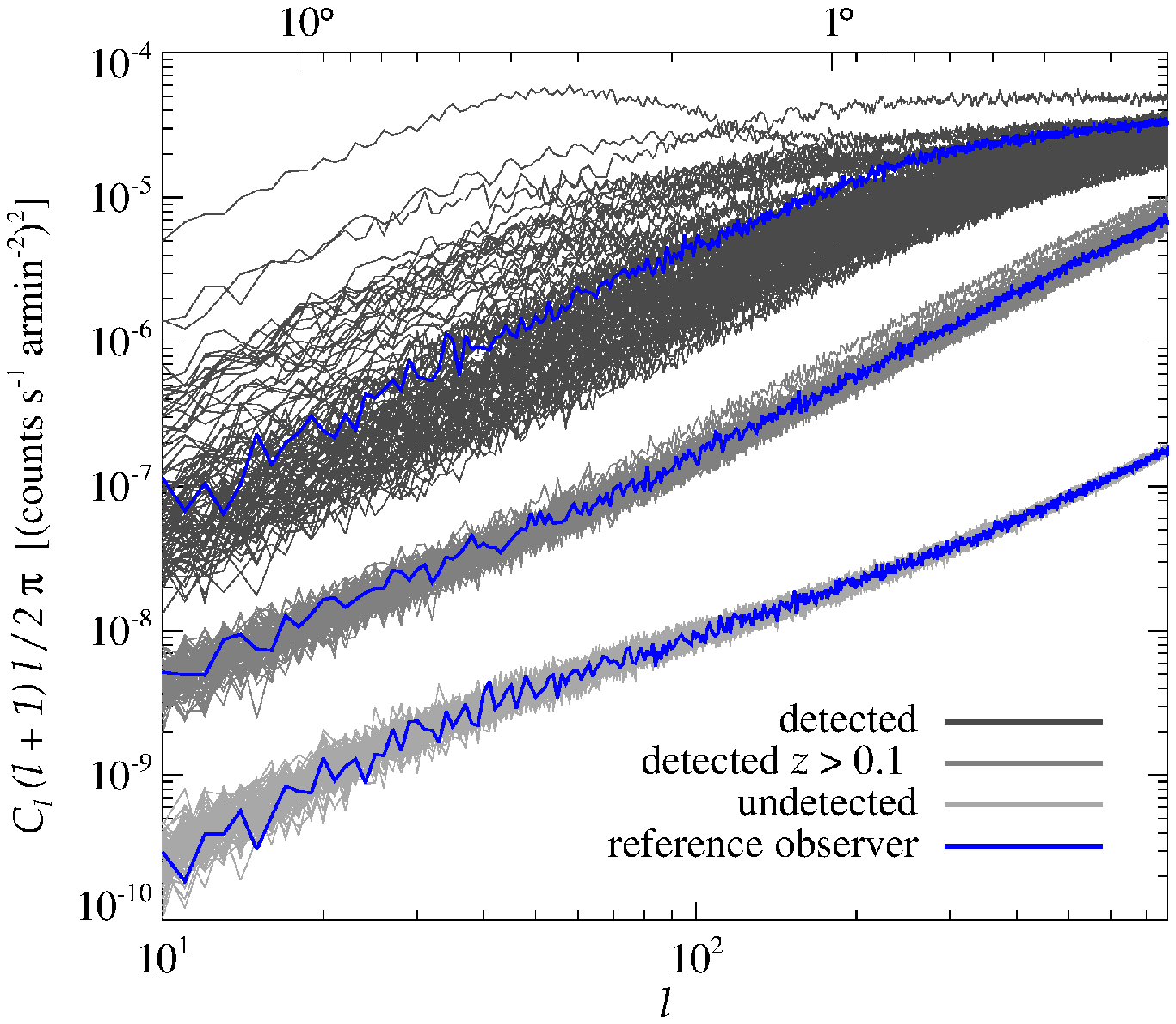}
\caption{\label{fig:cl_100} Variance among 100 random observers, for our baseline light-cone BigMDPL-LC with $1-b=0.8$, on the prediction for the angular power spectra of the eROSITA all-sky survey cluster population. The blue line is our reference observer. From top to bottom, the set of lines represent the eROSITA-detected clusters, the eROSITA-detected clusters with $z>0.1$, and the eROSITA-undetected population. These angular power spectra are obtained from HEALPix maps with a lower resolution with respect to what is shown in Figure~\ref{fig:cl} which results in a lower maximum $l$. See main text for details. The top axis indicates the angular scale as $180^{\circ}/l$.}
\end{figure}

The eROSITA all-sky survey has the advantage over the XBOOTES results, discussed in the previous section, of not being affected by sample variance, i.e., by the variance of the galaxy cluster samples in different small sky patches. It is, however, affected by cosmic variance that in the present paper we partially assessed by using 100 random observers from our baseline BigMDPL-LC with $1-b=0.8$. The result of this variance is shown in Figure~\ref{fig:cl_100} up to a lower maximum $l$ with respect to Figure~\ref{fig:cl} as these angular power spectra are calculated from HEALPix maps with $N_\mathrm{side} = 512$ (corresponding to a resolution of about $410$~arcsec) in order to speed up calculations. As we normalise the angular power spectra with the pixel area, i.e., the arcmin$^{-2}$ in the units of the y-axes, the resulting signals coincide with what is obtained using a higher resolution in the shown $l$-range.

The blue line in Figure~\ref{fig:cl_100} is our reference observer in the BigMDPL-LC with $1-b=0.8$ that nicely represents the average over the 100 random observers. When looking at the top set of lines for the eROSITA-detected population, the dominant effect of the nearby larger and brighter clusters in generating a huge variance on the angular power spectra is clear. In fact, this is also clearly shown by the all-sky maps in Figures~\ref{fig:maps_2} and \ref{fig:maps_3}. Therefore, in order to fruitfully use the eROSITA-detected clusters in such analyses it will be necessary to exclude the closest and brightest objects. The second, from the top, set of lines shows the reduction in variance among the 100 random observers for the eROSITA-detected population of clusters when excluding all objects at $z \leq 0.1$. Indeed, the variance is dramatically reduced and is already smaller than the differences among different models and cosmologies discussed above and shown in the top panels of Figure~\ref{fig:cl}. Excluding only clusters with $z \leq 0.05$ would also limit the huge variance seen in the top set of lines, while still showing a mildly larger variance than when excluding all objects at $z \leq 0.1$. Finally, the bottom set of lines in Figure~\ref{fig:cl_100} shows that the variance among the 100 random observers for the eROSITA-undetected population is much smaller than the previous case for $l \gtrsim 10^2$, and it is almost negligible at the largest multipoles (smallest) scales considered here. 

The results discussed in this section, and in the previous one, clearly show the potential of the angular power spectrum analysis of the eROSITA all-sky survey not only for doing cosmology (e.g., \citealp{2012MNRAS.422...44P}) but also to study the ICM - the normalisation and slope of X-ray scaling relations of clusters and the shape and clumpiness of the gas profile, particularly in the outskirts - and the composition of the cosmic X-ray background, and indicate that redshift tomography can be particularly important to limit the irreducible uncertainty due to cosmic variance.

%%%%%%%%%%%%%%%%%%%%%%%%%%%%%%%%%%%%%%%%%%%%%%%%%%%%%%%%%%%%%%%%%%%
%%%%%%%%%%%%%%%%%%%%%%%%%%%%%%%%%%%%%%%%%%%%%%%%%%%%%%%%%%%%%%%%%%%
\section{Limitations and future plans}
\label{sec:6}

As already discussed in the Introduction, our approach is advantageous with respect to generating a similar set of light-cones using hydrodynamical simulations mainly because of the prohibitive computational time that would be needed. In this context, alternative semi-analytical models to implement the baryonic physics on top of dark-matter-only halos do exist. While we adopted a phenomenological approach constructing our ICM using state-of-the-art observations, other models have a more theoretical-oriented approach: they build the ICM from first principles and then calibrate the model to match observations (e.g., \citealp{2010ApJ...725.1452S,2017ApJ...837..124F}). Such models, while implying a more elaborate procedure, permit a direct handling of some of the underlying fundamental physical parameters (as, e.g., the adiabatic index of the ICM). In the future we will investigate also this kind of models.

We can identify three categories for the limitations affecting our light-cones: limitations related to the used N-body simulations, limitations related to how the phenomenological ICM model is built, and limitations related to how certain quantities have been calculated.

The main limitations related to the used N-body simulations are two. First, we did not have enough redshift snapshots to accurately sample all the relevant redshift space where clusters of galaxies live. While we cover the most important part, we lack the snapshots to properly cover high redshift $z>1$. Second, our attempt to assess cosmic variance by generating 100 light-cones for 100 random observers in the BigMDPL simulation box is incomplete as we are always using the same simulation. We plan to overcome both problems with the next step in our endeavour where we will generate a thousand N-body simulations, tuned to the mass range more relevant for the eROSITA galaxy clusters, by using the new PPM-GLAM code \citep{2017arXiv170105690K}. Part of these simulations have already been done at the SURFsara Dutch supercomputing centre.

The main limitations of our phenomenological modelling are the assumption of spherical symmetry of the profiles, which, however, is pointless to abandon in the context of large mock catalogues, and the limited variation in profiles, i.e., the presence of only four dynamical categories for our clusters. There are a number of additional possibilities that could be investigated in our approach, including, e.g., different prescriptions for the definition of the $X_\mathrm{off}$ threshold values, the bias value, the redshift evolution, the temperature-mass relation, the metallicity value, and the inclusion of some break to recover the steeper luminosity-mass relation of galaxy groups with respect to clusters recently found by \cite{2017MNRAS.469.3738S}. In particular, the metal abundances in the ICM have complex radial profiles peaking the the clusters' centre (see, e.g., \citealp{2017A&A...603A..80M} and references therein): a higher/lower metallicity value with respect to our standard choice of a constant value of 0.3 solar would imply a higher/lower X-ray emission. On a related note, the already mentioned gas clumpiness (see, e.g., \citealp{2013MNRAS.429..799V,2015MNRAS.447.2198E}) would also boost the total X-ray emission with respect to a smooth gas profile.
These points can be investigated in future releases. Let us also note two shortcomings of our approach for which there is no simple solution: the scatter is added independently on the pressure and on the temperature, while, in reality, these two quantities should be related, and, as discussed in Section~\ref{sec:3}, the hydrostatic bias and temperature dependencies remain encoded in $f_\mathrm{gas}=M_\mathrm{gas,HE}/M_{500, \mathrm{HE}}$. 

The main limitations in some of the calculated quantities are related to how we computed the eROSITA counts. As already explained in Section~\ref{sec:2.2}, we followed the approach of \cite{2012MNRAS.422...44P} and adopted a number of simplifications with respect to reality. In particular, we assumed a uniform exposure of the sky by the eROSITA telescope, a uniform value for the hydrogen column density for the photoelectric absorption, and we did not consider any background. Additionally, we also adopted a simple detection criteria by defining detected by eROSITA clusters with $\geq50$ counts (within a given integration radius $R_{500}$, $2.5 \times R_{500}$, or $5 \times R_{500}$) in 1.6~ks of observations. While this turns out to be a good approximation for the purposes of the present work \citep{2018arXiv180608652C}, it is worth mentioning, as also noted in Section~\ref{sec:5.2}, that different ICM profiles can result in changing the detection likelihood: i.e., a much flatter X-ray surface brightness profile could make harder the detection because more background counts would slip in the relevant cluster regions. However, the eROSITA selection function corresponding to such alternative profile choices need to be studied in detail, as done in \cite{2018arXiv180608652C} for an average $\beta$-model profile, before we can make an informed decision on alternative detection criteria. At any rate, these are all aspects which can affect our results and will be investigated in future releases. Another point worth mentioning here is that our all-sky angular power spectra for eROSITA are theoretical predictions that do not account for effects related to the imperfect removal of point sources or of the Galactic background.

Some less important simplifications were done in calculating some of the light-cone entries to speed up the computation. For example, we added a posteriori the expected scatter on the X-ray luminosity and SZ effect on the light-cones containing masses down to $M_{500}\geq1\times10^{13}$~$h_\mathrm{cosmo}^{-1}$~M$_\mathrm{sun}$ as we use an interpolation procedure to obtain these quantities in that case (see Section~\ref{sec:2.2}). Additionally, some light-cones contain a limited number of entries for economy of disk space as in the case of the 100 random-observer light-cones. Such second order simplifications and choices have all be mentioned throughout the text and have been driven by the scientific scope of the present work and by the need of making available a general enough product for the community. Other choices can be made for specific needs and we encourage the possible light-cone user to contact us in case our choices do not suite his/her scientific case.

%%%%%%%%%%%%%%%%%%%%%%%%%%%%%%%%%%%%%%%%%%%%%%%%%%%%%%%%%%%%%%%%%%%
%%%%%%%%%%%%%%%%%%%%%%%%%%%%%%%%%%%%%%%%%%%%%%%%%%%%%%%%%%%%%%%%%%%
\section{Summary}
\label{sec:7}

In this work we presented the first release of MultiDark-Clusters: i.e., a set of galaxy cluster mock light-cones constructed from the dark-matter only MultiDark simulations that reproduces current state-of-the-art X-ray and SZ observations. We presented the methodology used to generate the light-cones from the simulation snapshots and to implement the baryonic physics onto the MultiDark halos. We then discussed two applications: predictions for the eROSITA cluster population and for its power spectrum, particularly discussing the potential of the latter not only for cosmological studies but also to study the ICM and the composition of the cosmic X-ray background. The main results of this paper are the following. 

\begin{itemize}
\item We developed a phenomenological model -- starting from the \emph{Planck} pressure profiles and including four different dynamical states for clusters -- to implement ICM properties onto dark-matter-only halos. We showed that our phenomenological model well reproduces state-of-the-art X-ray and SZ observations of clusters of galaxies.\\
\item We constructed a set of galaxy cluster mock light-cones from the BigMD27 and BigDMPL MultiDark simulations with a volume of $(2.5~h^{-1}~\mathrm{Gpc})^{3}$ and with WMAP-7 and \emph{Planck} year-1 cosmologies, respectively. We make freely available online on Skies \& Universes (http://www.skiesanduniverses.org; \citealp{2017arXiv171101453K}) more than 400~GB of light-cone catalogues. We discussed current limitations and plans for future releases in the last section of the paper.\\
\item We showed predictions for the cluster number counts for the eROSITA all-sky survey. In particular, we discussed the differences when adopting different phenomenological models, different underlying cosmologies, and the effect of cosmic variance that we partially investigated by using 100 random-observer light-cones constructed from the BigDMPL simulation.\\
\item We compared the power spectrum of our light-cones with current state-of-the-art measurements of the X-ray surface brightness fluctuations of the XBOOTES field \citep{2017MNRAS.466.3035K,2018MNRAS.473.4653K}. We argued for the need to overcome sample variance with all-sky surveys as the eROSITA one in order to fully exploit the potential of such an approach to study the ICM of clusters and the composition of the cosmic X-ray background.\\ 
\item We showed the full potential of the power-spectrum technique by discussing predictions for the angular power spectrum of the eROSITA all-sky detected and undetected cluster population. We discussed again the differences among different models and cosmologies, and we investigated cosmic variance by using 100 random-observer light-cones. We noted, in particular, that redshift tomography with the eROSITA cluster population will be fundamental to limit the irreducible uncertainty due to cosmic variance.
\end{itemize}

%%%%%%%%%%%%%%%%%%%%%%%%%%%%%%%%%%%%%%%%%%%%%%%%%%%%%%%%%%%%%%%%%%%
%%%%%%%%%%%%%%%%%%%%%%%%%%%%%%%%%%%%%%%%%%%%%%%%%%%%%%%%%%%%%%%%%%%
\section*{Acknowledgments}
We would like to thank the anonymous referee for the useful suggestions. FZ would additionally like to thank the anonymous referee 
for comparing the early version of the manuscript's footnotes to those of David Foster Wallace; FZ considers this a compliment and
surely the highest point in his career. We would like to thank A.~Kolodzig, K.~Borm, J.~Comparat, A.~Merloni and D.~Eckert for useful discussions and comments.
The BigMD simulation suite have been performed in the Supermuc supercomputer at LRZ using time granted by PRACE.
FZ acknowledges the supported by the Netherlands Organization for Scientific Research (NWO) through a Veni grant.
MF acknowledges the supported by NWO through a Vidi grant assigned to Shin'ichiro Ando. 
F.~Prada and AK acknowledge support from the Spanish MINECO grant AYA2014-606Y1-C2-1-P.
THR and F.~Pacaud acknowledge support by the German Research Association (DFG) through the Transregional Collaborative Research Centre TRR33 The Dark Universe (project B18) as well as by the German Aerospace Agency (DLR) with funds from the Ministry of Economy and Technology (BMWi) through grant 50 OR 1514.

%%%%%%%%%%%%%%%%%%%%%%%%%%%%%%%%%%%%%%%%%%%%%%%%%%%%%%%%%%%%%%%%%%%
%%%%%%%%%%%%%%%%%%%%%%%%%%%%%%%%%%%%%%%%%%%%%%%%%%%%%%%%%%%%%%%%%%%
\bibliographystyle{mn2e}
\bibliography{bib_file}

%%%%%%%%%%%%%%%%%%%%%%%%%%%%%%%%%%%%%%%%%%%%%%%%%%%%%%%%%%%%%%%%%%%
%%%%%%%%%%%%%%%%%%%%%%%%%%%%%%%%%%%%%%%%%%%%%%%%%%%%%%%%%%%%%%%%%%%
\begin{appendix}

\section{Description of the MultiDark-Clusters mock light-cones available on-line}
\label{app:A}

One of the goals of our endeavour is to make publicly available the produced mock light-cones. We start by making available on-line at Skies \& Universes (http://www.skiesanduniverses.org; \citealp{2017arXiv171101453K}) all the 124 light-cones produced and used in this work, for an amount of data of more than 400 GB, that we dubbed MultiDark-Clusters. 
We refer the reader to Section~\ref{sec:2} for all the details on how the light-cones have been generated. However, we want to stress again here that the light-cones are accurate within a limited redshift range: below $z \leq 0.94$ for BigMDPL and only within $ 0.16 \leq z \leq 0.86$ for BigMD27.\\ 

The produced light-cones are divided in three categories as follows.
\vspace{0.1cm}
\begin{enumerate}
\item The light-cones obtained following the procedure described by items (i)-(v) in Section~\ref{sec:2.2} that include clusters with masses $M_{500}\geq5\times10^{13}$~$h_\mathrm{cosmo}^{-1}$~M$_\mathrm{sun}$. There are 10 of these catalogues, 5 for the \emph{Planck} cosmology and 5 for the WMAP-7 cosmology. They are dubbed in the following way. BigMDPL$\_$lightcone$\_$5e13$\_$Mantz2010 refers to the \emph{Planck} cosmology, adopts the $M_{500}-T$ relation of \cite{2010MNRAS.406.1773M}, and our reference value for $1-b=0.8$. If it adopts the \cite{2016MNRAS.463.3582M} $M_{500}-T$ relation, it is dubbed $\_$Mantz2010, and if it adopts a different $1-b$ value, this is added at the end of the name file as $\_$bias$X$ with $X=1-b=0.6, 1$ or $1.2$. The light-cones for the WMAP-7 cosmology use the same nomenclature but start with BigMD27.\\
\item The light-cones obtained following the procedure described in item (vi) of Section~\ref{sec:2.2} that include clusters with masses $M_{500}\geq1\times10^{13}$~$h_\mathrm{cosmo}^{-1}$~M$_\mathrm{sun}$ and, by construction, contain a limited set of entries with respect to the ones of point (i). There are 14 of these catalogues, 5 for each cosmology, \emph{Planck} and WMAP-7, for the different phenomenological models as above, and 4 additional catalogues for the \emph{Planck} cosmology (with the $M_{500}-T$ relation of \cite{2010MNRAS.406.1773M} and our reference value for $1-b=0.8$) obtained adopting a different limiting integration radius with respect to $R_{500}$, i.e., $2.5 \times R_{500}$ and $5 \times R_{500}$, and adopting $5 \times R_{500}$ but with two different underlying pressure profiles dubbed $P_\mathrm{upper}$ and $P_\mathrm{lower}$ (see Section~\ref{sec:5} and Figure~\ref{fig:P_envelope}). The nomenclature of the files is as for the ones of point (i) with the addendum of inter (standing for interpolation).
For example, BigMD27$\_$lightcone$\_$inter$\_$1e13$\_$Mantz2010. The last 4 described catalogues are clearly identified by an addendum at the end of the file names referring to the adopted radius and pressure.\footnote{If one would like to extend the projected profiles to larger radii, or radii not currently available on-line, we underline that this cannot be done by extrapolating the provided profiles out to the desired radial extension as the shape would be wrong due to the radial integral limit in Eq.~(5) and (6). In order to have the projected profiles out to larger radii, one would have to re-do the projection starting from the radial profiles.}\\
\item The 100 light-cones for the 100 random observers for the case of the \emph{Planck} cosmology adopting the $M_{500}-T$ relation of \cite{2010MNRAS.406.1773M} and our reference value for $1-b=0.8$. These are obtained following the procedure described in item (vi) of Section~\ref{sec:2.2} and include clusters with masses $M_{500}\geq1\times10^{13}$~$h_\mathrm{cosmo}^{-1}$~M$_\mathrm{sun}$, as the light-cones of point (ii). However, they contain a reduced number of entries, more relevant for the X-ray application of this paper, with respect to the ones of point (ii).
\end{enumerate}

\vspace{0.1cm}
The light-cones belonging to the category (i) above contain the following entries for each galaxy cluster.
\vspace{0.1cm}
\begin{enumerate}
\item[(1)] $D$ [$h_{\mathrm{cosmo}}^{-1}$ Mpc] = luminosity distance.\\
\item[(2)] $z$ = redshift.\\
\item[(3)] $l$ [deg] = Galactic longitude.\\
\item[(4)] $b$ [deg] = Galactic latitude.\\
\item[(5)] $M_{200}$ [$h_{\mathrm{cosmo}}^{-1}$ M$_{\odot}$] = mass defined with respect to 200 times the critical density of the Universe.\\
\item[(6)] $M_{500}$ [$h_{\mathrm{cosmo}}^{-1}$ M$_{\odot}$] = mass defined with respect to 500 times the critical density of the Universe.\\
\item[(7)] $R_{500}$ [$h_{\mathrm{cosmo}}^{-1}$ Mpc] = radius corresponding to $M_{500}$.\\
\item[(8)] $T_{drop}$ [\#] = central temperature drop (0.4, 0.6, 0.8, 1) that defines the type of cluster.\\
\item[(9)] $P_{500}$ [$h_{\mathrm{cosmo}}^{1/2}$ keV cm$^{-3}$] = pressure normalisation defined with respect to $R_{500}$.\\
\item[(10)] $T_\mathrm{Mantz}$ [keV] = temperature from the centrally-excised $M_{500}-T$ relation of Mantz et al.~(2010) or Mantz et al.~(2016) depending on the catalogue.\\
\item[(11)] $T_{500}$  [keV] = final $R_{500}$-volume-averaged temperature (not centrally-excised) used as input for the XSPEC apec model to obtain volume-integrated fluxes, luminosities and counts.\\
\item[(12)] APEC$_\mathrm{norm}$ [cm$^{-5}$] = XSPEC apec normalisation within $R_{500}$.\\
\item[(13)] $Y_\mathrm{SZ}$ [$h_{\mathrm{cosmo}}^{-2.5}$ Mpc$^{2}$] = Sunyaev-Zel'dovich signal within $R_{500}$.\\
\item[(14)] $Y_\mathrm{SZ,HE}$ [$h_{\mathrm{cosmo}}^{-2.5}$ Mpc$^{2}$] = Sunyaev-Zel'dovich signal within $R_{500, \mathrm{HE}}$ which refers to the hydrostatic-biased mass.\\
\item[(15)] $Y_\mathrm{X}$ [$h_{\mathrm{cosmo}}^{-2.5}$ M$_\odot$ keV] = M$_\mathrm{gas} \times T_{500, \mathrm{HE}}$ within $R_{500, \mathrm{HE}}$ for comparison with Vikhlinin et al.~(2009).\\
\item[(16)] $M_\mathrm{gas}$ [$h_{\mathrm{cosmo}}^{-2.5}$ M$_\odot$] = gas mass calculated from the gas profile integrated within $R_{500, \mathrm{HE}}$.\\
\item[(17)] $F_{0.1-2.4}$ [erg cm$^{-2}$ s$^{-1}$] = XSPEC apec observer-frame unabsorbed flux within $R_{500}$ and $0.1-2.4$ keV energy range (metallicity is fixed to 0.3).\\
\item[(18)] $F_{0.5-2}$ [erg cm$^{-2}$ s$^{-1}$] = as above but for the $0.5-2$ keV energy range.\\
\item[(19)] $L_{0.1-2.4}$ [erg s$^{-1}$] = XSPEC APEC rest-frame unabsorbed luminosity within $R_{500}$ and $0.1-2.4$ keV energy range.\\
\item[(20)] $L_{0.5-2}$ [erg s$^{-1}$] = as above but for the $0.5-2$ keV energy range.\\
\item[(21)] $L_\mathrm{bol}$ [erg s$^{-1}$] = as above but bolometric in the the $0.01-100$ keV energy range.\\
\item[(22)] count rate [ph s$^{-1}$] = observer-frame eROSITA count rate (including absorption) within $R_{500}$ and in the $0.5-2$ keV energy range obtained as in Pillepich et al.~(2012) without Poissonian noise.\\
\item[(23) to (32)]$F_{0.5-2,\mathrm{proj}}^{0-9}$ [erg cm$^{-2}$ s$^{-1}$] = observer-frame unabsorbed fluxes in the $0.5-2$ keV energy range of galaxy clusters projected onto the sky corresponding to 10 spherical shells at $r_{i=0-9} = (\Delta r \times i) + \Delta r/2$ with thickness $\Delta r = R_{500}/10$, used to describe the X-ray profile of each cluster, if summed return F$_{0.5-2}$ (entry 18).\\
\item[(33) to (42)] $Y_\mathrm{SZ,proj}^{0-9}$ [$h_{\mathrm{cosmo}}^{-2.5}$ Mpc$^{2}$] = same as above but for the Sunyaev-Zel'dovich signal, if summed return Y$_\mathrm{SZ}$ (entry 13).
\end{enumerate} 

\vspace{0.1cm}
 
The light-cones belonging to the category (ii)  above and dubbed inter contain only columns 1--to--8, 13 and 17--to--42 for each galaxy cluster For the light-cones adopting a different integration radius, i.e., $2.5 \times R_{500}$ and $5 \times R_{500}$, the entries 13 and 17--to--22 refer to the chosen integration radius rather than to $R_{500}$. For the light-cone adopting $2.5 \times R_{500}$ as integration radius, the profiles entries 23--to--42 refer to $\Delta r = 2.5 \times R_{500}/10$. For the light-cones adopting $5 \times R_{500}$ as integration radius, the profiles entries 23--to--42 refer only to the X-ray quantity $F_{0.5-2,\mathrm{proj}}^{0-19}$ with $r_{i=0-19} = (\Delta r \times i) + \Delta r/2$ and thickness $\Delta r = 5 \times R_{500}/20$.\\ 

The 100 light-cones for the random observers belonging of category (iii) above contain only columns 2--to--4, 6, 8,  17--to--20 and 22. These are dubbed with numbers from 0 to 99. The observer corresponding to the BigMDPL reference observer of the previous categories, (i) and (ii), of light-cones is number 17.

\end{appendix}

%%%%%%%%%%%%%%%%%%%%%%%%%%%%%%%%%%%%%%%%%%%%%%%%%%%%%%%%%%%%%%%%%%
%%%%%%%%%%%%%%%%%%%%%%%%%%%%%%%%%%%%%%%%%%%%%%%%%%%%%%%%%%%%%%%%%%
\label{lastpage}

\end{document}